 \documentclass[preprint,review,12pt,sort&compress]{elsarticle}
\usepackage[margin=1.18 in]{geometry}
\usepackage{multirow}
\usepackage{adjustbox}
\usepackage{graphicx}
\usepackage{gensymb}
\usepackage{xcolor}
\definecolor{UW}{RGB}{64, 38, 96}
\usepackage{verbatim}
\newcommand{\B}{Ba\v zant}  

 \newcommand{\bc}{\begin{center}}
 \newcommand{\ec}{\end{center}}
                   \newcommand{\bfr}{\begin{flushright}}
                   \newcommand{\efr}{\end{flushright}}
   \newcommand{\ii}{\item}
     \newcommand{\be}{\begin{enumerate}}
     \newcommand{\ee}{\end{enumerate}}
        \newcommand{\bi}{\begin{itemize}}
        \newcommand{\ei}{\end{itemize}}
            \newcommand{\bd}{\begin{description}}
            \newcommand{\ed}{\end{description}}
                \newcommand{\beq}{\begin{equation}}
                \newcommand{\eeq}{\end{equation}}
                  \newcommand{\bea}{\begin{eqnarray}}
                  \newcommand{\eea}{\end{eqnarray}}

      \newcommand{\bfi}{\begin{figure}}
      \newcommand{\efi}{\end{figure}}
\newcommand{\bay}{\begin{array}{l}}
\newcommand{\eay}{\end{array}}
            
\usepackage{amsmath}
\usepackage{breqn}
\usepackage{textcomp}

\usepackage{amssymb}

\journal{Composite Structures}

\begin{document}

\begin{titlepage}

\clearpage\thispagestyle{empty}



\noindent

\hrulefill

\begin{figure}[h!]

\centering

\includegraphics[width=1.5 in]{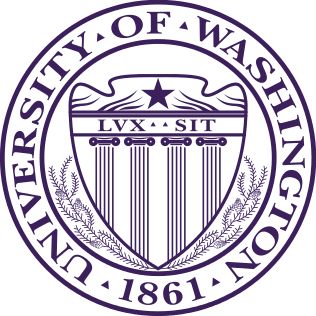}

\end{figure}


\begin{center}

{\color{UW}{

{\bf A\&A Program in Structures} \\ [0.1in]

William E. Boeing Department of Aeronautics and Astronautics \\ [0.1in]

University of Washington \\ [0.1in]

Seattle, Washington 98195, USA

}

}

\end{center} 

\hrulefill \\ \vskip 2mm

\vskip 0.5in

\begin{center}

{\large {\bf Fracturing Behavior and Size Effect of Discontinuous Fiber Composite Structures with Different Platelet Sizes}}\\[0.5in]

{\large {\sc Seunghyun Ko, Jinkyu Yang, Mark E. Tuttle, Marco Salviato}}\\[0.75in]

{\sf \bf INTERNAL REPORT No. 18-12/04E}\\[0.75in]

\end{center}

\noindent {\footnotesize {{\em Submitted to Composite Structures \hfill December 2018} }}

\end{titlepage}

\begin{frontmatter}



\title{Fracturing Behavior and Size Effect of Discontinuous Fiber Composite Structures with Different Platelet Sizes}

\author[label1]{Seunghyun Ko}
\author[label1]{Jinkyu Yang}
\author[label2]{Mark E. Tuttle}
\author[label1]{Marco Salviato \corref{cor1}}
 \address[label1]{William E. Boeing Department of Aeronautics and Astronautics, University of Washington, Seattle, WA 98195, USA}
 \address[label2]{Department of Mechanical Engineering, University of Washington, Seattle, WA 98195, USA}

\cortext[cor1]{Corresponding Author, \ead{salviato@aa.washington.edu}}

\begin{abstract}
\linespread{1}\selectfont

This study investigates the mode I intra-laminar fracture and size effect in Discontinuous Fiber Composites (DFCs). Towards this goal, the results of fracture tests on geometrically-scaled Single Edge Notch Tension (SENT) specimens are presented and critically discussed for three platelet sizes.

The results clearly show a decrease in nominal strength as the specimen size increases. This effect becomes more important as the structure size increases. It is found that, when the specimen is sufficiently large, the structural strength scales according to Linear Elastic Fracture Mechanics (LEFM) and the failure occurs in a very brittle way. In contrast, small specimens exhibit a more pronounced pseudo-ductility with a limited scaling effect and a significant deviation from LEFM.  

To characterize the fracture energy and the effective length of the fracture process zone, an approach combining equivalent fracture mechanics and stochastic finite element modeling is proposed. The model accounts for the complex random mesostructure of the material by modeling the platelets explicitly. Thanks to this theoretical framework, the mode I fracture energy of DFCs is estimated for the first time and it is shown to depend significantly on the platelet size.
 In particular, the fracture energy is shown to increase linearly with the platelet size in the range investigated in this work.  

Another important conclusion of this work is that, compared to traditional unidirectional composites, DFC structures exhibit higher pseudo-ductility and their strength is, by far, less sensitive to notches, defects and cracks. However, this aspect can be used advantageously in structural design only upon the condition that proper certification guidelines acknowledging the more pronounced quasibrittleness of DFCs is formulated. The size effect analysis presented in this work represents a first step in this direction as it allows the assessment of the severity of a defect or notch in DFCs. 
\end{abstract}


\end{frontmatter}




\section{Introduction}
Composites reinforced by randomly oriented platelets or short fibers, generally called Discontinuous Fiber Composites (DFCs), offer several advantages over traditional unidirectional composites. Not only do they feature a relatively pseudo-ductile behavior \cite{Io2006, Fera2009a, Fera2009b, Wan2016, Shin2016, Sel2015}, but they also enable the manufacturing of parts in complex shapes without the need for machining or the use of adhesives \cite{Quantum, Hexcel, Toyota}. This is thanks to significant formability, unmatched by traditional unidirectional carbon fiber composites, which makes the use of compression molding a highly viable option even for very complex geometries \cite{Boeing, ForgedComposite, Tencate}. This enables almost net-shape designs with minimum waste of materials \cite{Tencate}. Further, the platelet-based geometry opens new avenues for recycling uncured prepreg materials \cite{Nila2015, Jin2017} and developing hybrid laminates with continuous fibers to achieve unprecedented mechanical properties and formability \cite{Vis2017}. These characteristics broaden the use of DFCs for applications that have been typical of light alloys, such as secondary structural components for aerospace \cite{Boeing,Hexcel,Tencate}, body frames of terrestrial vehicles \cite{ForgedComposite, Hexcel, Toyota, Quantum}, composite brackets, suspension arms and interiors \cite{ForgedComposite} and crash absorbers \cite{Quantum, Hexcel, ForgedComposite}. 

Considering the remarkable properties of DFCs, it is not surprising that the scientific and industrial communities devoted significant efforts to understanding their mechanical behavior and developing proper design guidelines. Since the pioneering work of Halpin and Pagano \cite{HalPag69}, significant progress have been made on the experimental and computational characterization of the DFC mesostructure \cite{Wan2018, Denos2018a, Denos2018b, Sommer2017}, and the understanding of the influence of the platelet morphology on the elastic properties and strength of DFCs \cite{Io2006, Fera2009a, Fera2009b, Wan2016, Shin2016, Sel2015, Pipes2018, Pipes2019}. However, while a large bulk of data on the mechanical properties of DFCs is available already, an aspect overlooked in the literature is the fracturing behavior and its scaling effect. This is a serious issue since the design of complex-shaped DFC components featuring holes, notches, and other stress raisers requires a thorough understanding of the fracturing process and its size effect in particular. 

This is also challenging since the DFCs are the quasibrittle structures, i.e. structures made of complex, heterogeneous materials with non-negligible inhomogeneities. For typical engineering applications, materials that lead to a quasibrittle behavior include e.g. concrete \cite{Baz1984, Baz1990, Baz1998}, composites \cite{Baz1996, Sal2016a, Sal2016b}, and nanocomposites \cite{Sal2011, Sal2013, Yao2018a, Yao2018b}. In quasibrittle structures, the size of the non-linear Fracture Process Zone (FPZ) occurring in the presence of a large stress-free crack is usually not negligible  \cite{Baz1984, Baz1990, Baz1996, Baz1998, Sal2016a, Yao2018a, Yao2018b, Cusatis2018}. Particularly in DFCs, the stress field along the FPZ is nonuniform and decreases with crack opening due to discontinuous cracking, delamination and frictional pullout of platelets \cite{Ko2018a, Ko2018b}. As a consequence, the fracturing behavior and, most importantly, the energetic size effect associated with the given structural geometry, cannot be described by means of the classical Linear Elastic Fracture Mechanics (LEFM). To capture the effects of a finite, non-negligible FPZ, the introduction of a characteristic (finite) length scale related to the fracture energy and the strength of the material is essential \cite{Baz1984, Baz1990, Baz1996, Baz1998, Sal2016a, Yao2018a, Yao2018b, Cusatis2018}.


Another aspect that makes the characterization of the fracturing behavior of DFCs particularly challenging is the significant role played by the platelet morphology, especially the platelet size and its spatial random distribution throughout the structure. In fact, the random orientation distribution of platelets leads to significant spatial variability of the mechanical properties as well as of the local stress field \cite{Sel2015,Ko2018b}. In particular, the final failure caused by damage localization is often initiated by the presence of weak spots rather than stress raisers such as cracks or notches. The location of these weak spots is dictated by the spatial randomness of both local material resistance and applied stress field.  

In consideration of the foregoing knowledge gaps, this study presents an investigation of the intra-laminar fracture and size effect in DFCs for three platelet sizes ($75\times12$, $50\times8$, and $25\times4$ mm). The size effect on the structural strength of geometrically-scaled Single Edge Notch Tension (SENT) specimens is characterized for the first time, showing that neither stress-based failure criteria nor LEFM can solely capture the scaling effect. To capture the size effect and characterize the fracture energy, $G_f$, and the effective length of the fracture process zone, $c_f$, an approach combining equivalent fracture mechanics and stochastic finite element modeling is proposed in this study. The model accounts for the complex random mesostructure of the material by modeling the platelets explicitly. Thanks to this theoretical framework, the mode I fracture energy of DFCs is estimated for the first time and it is shown to depend linearly on the platelet size for the size range investigated in this work. It is found that the fracture energy of DFCs is much larger than several light alloys and always comparable to or larger than the quasi-isotropic laminate composite made from the same constituents. This aspect is significantly interesting considering the possible use of DFCs in the applications demanding crashworthiness \cite{Matsuo2017}. 

Finally, it is shown that structures made of DFCs are significantly less brittle compared to the ones made of traditional laminated composites. This aspect is particularly important for the formulation of certification guidelines for DFCs and the development of strategies for inspection and maintenance.

\section{Material preparation and test description}

\subsection{DFC manufacturing procedure}
To control the dimensions of the platelets precisely, an \textit{in-house} manufacturing process based on semi-automated manufacturing protocols were developed. The main steps of the processes used to manufacture the specimens can be summarized as follows \cite{Ko2018a, Ko2018b}:

\be  \setlength{\itemsep}{1.3mm}

\begin{figure}
 \centering
  \includegraphics[trim=0cm 1cm 0cm 0cm, clip=true,width = 1\textwidth, scale=1]{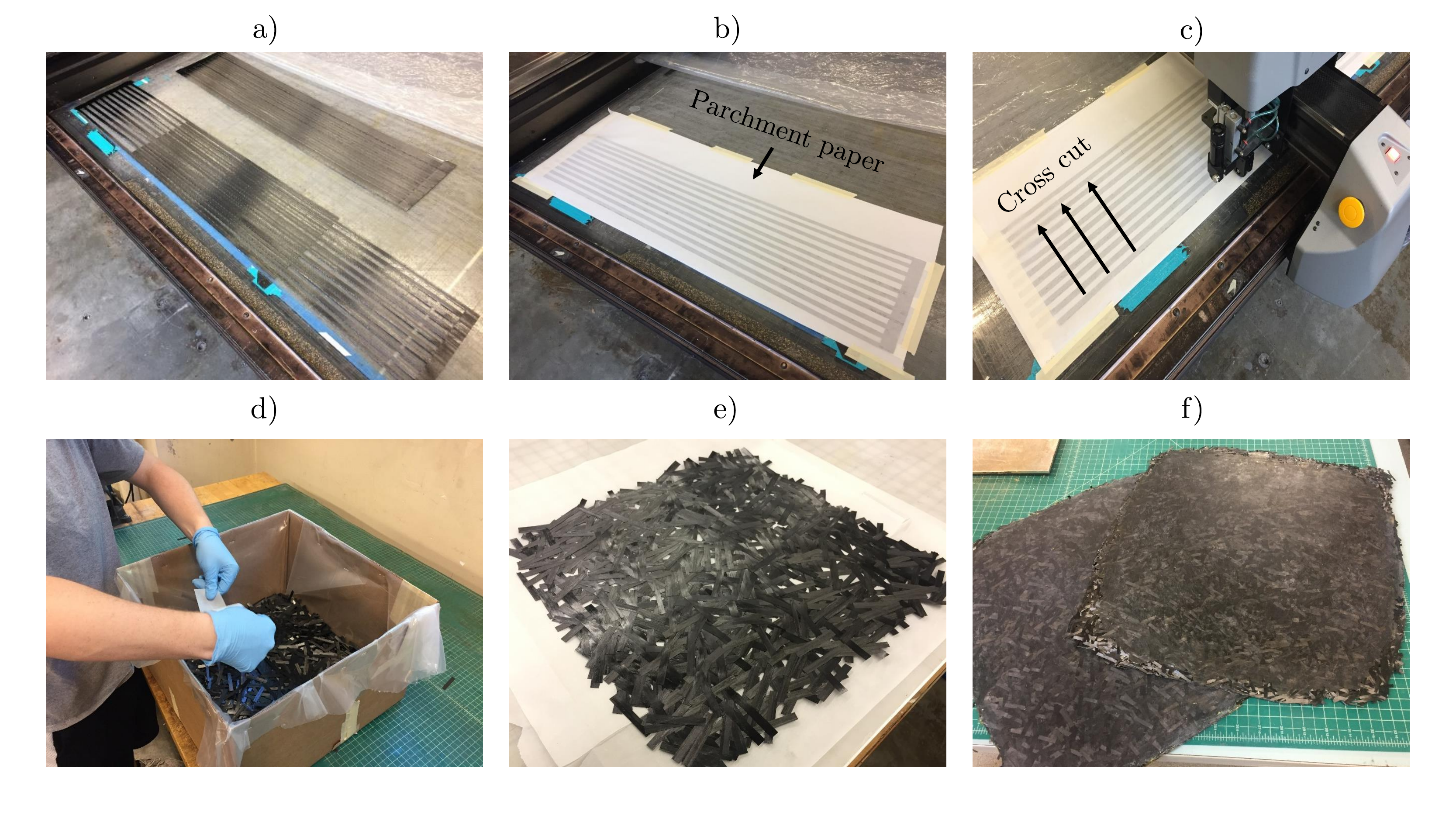} \caption{\label{f1} \sf In-house manufacturing process of the DFCs.} 
\end{figure}

\ii Sheets of Toray P$707$AG-$15$ with T$700$G-$12$K fiber reinforced prepreg were cut into two comb-shaped strips using a CNC fabric cutter (manufactured by Autometrix Advantage). Each strip was cut with a width corresponding to the desired platelet dimensions (Fig.~\ref{f1}a);

\ii The protective backing tape was removed. Both surfaces of the prepreg strips were covered using silicone coated parchment papers. The parchment papers acted as protective layers which could be easily removed after the final cuts (Fig.~\ref{f1}b);

\ii The prepreg strips were cross-cut to the desired length using the CNC fabric cutter. The result was a batch of platelets with very accurate dimensions (Fig.~\ref{f1}c);

\ii The platelets were manually shaken into a container in a random fashion to form a mat (Fig.~\ref{f1}d); 

\ii The mat stacking method proposed by Jin \emph{et al} \cite{Jin2017} was adapted to control the thickness of the plate. Eight mats of platelets, each weighing around $100$ g (see Fig.~\ref{f1}e), were used to build the plate with $3.3$ mm thickness;

\ii The plate was debulked under $100$ kPa for two hours. The debulking process removed air pockets trapped between the platelets and enhanced the uniform thickness throughout the plate;

\ii The DFC plate was transported to the hot press for the curing. The plate was cured at $270$ $\degree$F for two hours under $0.61$ MPa;

\ii The cured plate (see Fig.~\ref{f1}f) was trimmed $15$ mm from the edges to remove possible uncured spots. 
 
\ee

A thorough analysis of the plates confirmed the robustness of the proposed manufacturing approach. In fact, the volume fraction of voids was far below $2 \%$ for all the plates manufactured in this work whereas the thickness was provided with a tolerance of $\pm 3.31 \%$. 


\subsection{Specimen characteristics}
Previous studies \cite{Baz1984, Baz1996, Sal2016a} investigated the intra-laminar size effect on unidirectional and textile composites and successfully obtained the fracture energy. The present work extends the experimental procedure proposed by Salviato \emph{et al.} \cite{Sal2016a} to account for the peculiar characteristics of the DFCs.

\begin{figure}
 \centering
  \includegraphics[trim=2cm 1cm 2cm 0cm, clip=true,width = 1\textwidth, scale=1]{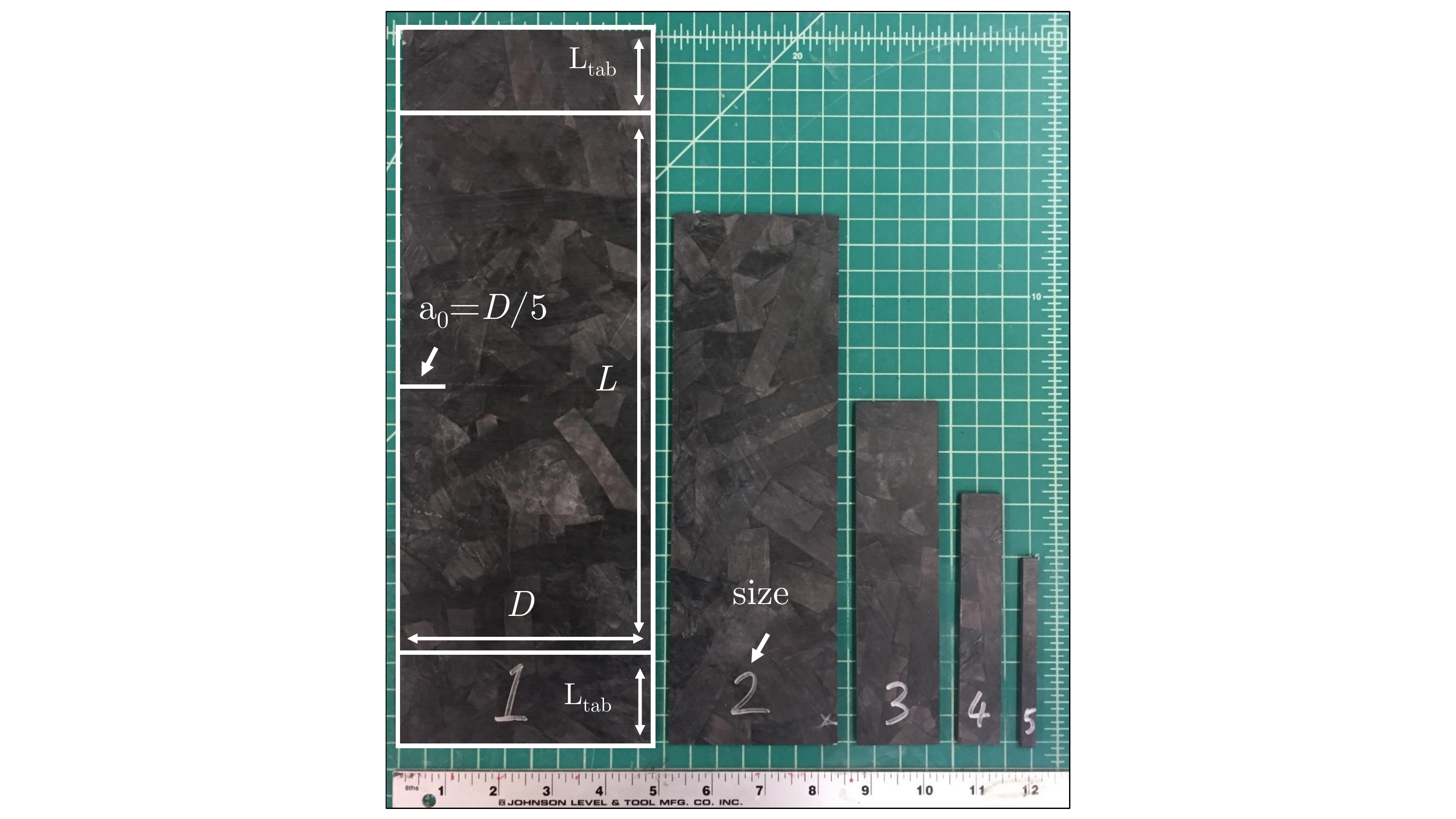} \caption{\label{f2} \sf Geometrically-scaled, Single Edge Notch Tension specimens investigated in this work.} 
\end{figure}

Fracture tests on geometrically-scaled Single Edge Notched Tension (SENT) specimens (see Fig.~\ref{f2}) of five different sizes were conducted. The ratio between size-$1$ (largest) and the size-$5$ (smallest) was $19:1$. The specimens were geometrically scaled except the thickness which was kept to a constant value of $3.3$ mm. Table~\ref{T1} summarizes the geometrical details of the specimens. Five different sizes of coupons were prepared for all three platelets dimensions, $75\times12$, $50\times8$, and $25\times4$ mm. These platelet dimensions were chosen in comparison with the $50\times8$ mm-sized platelets used for commercial aerospace DFC structures \cite{Hexcel}.


To create a sharp notch, a thin, diamond-coated, razor blade saw was used. The blade thickness was $0.2$ mm. The width of the notch was kept in a constant ratio relative to the structure width in this study ($a_0 = D/5$ in Fig.~\ref{f2}). A layer of white paint was sprayed followed by the black speckles to use Digital Image Correlation (DIC) technique. It is noteworthy that the displacement fields of the DFC specimens on the front and rear surfaces could be different because of the inhomogeneous characteristics of the material \cite{Joh2015}. However, a single surface DIC result was sufficient to observe the displacement field near the notch and to enable the accurate measurement of the displacement in the gauge area of the specimens and to characterize size-dependent fracture behavior in DFCs. 

\subsection{Testing}
A closed-loop, servo-hydraulic Instron 5585H with $200$ kN capacity was used for all the tests. The nominal strain rate was set to $0.2$ $\%$/min for all the specimen sizes investigated in this work. The load from the machine was recorded with a sampling frequency of $10$ Hz. To use the DIC technique, digital images were captured using a Nikon D$5600$ DSLR camera with Nikon AF micro $200$ mm and Sigma $135$ mm DG HSM lenses. The images were taken with a sampling rate of $1$ Hz.


\section{Experimental results}
\subsection{Load-displacement curves} 
The load-displacement curves of the size effect tests were analyzed based on the displacement field computed from DIC using the GOM Correlate software \cite{GOM}.. The nominal displacement was calculated by averaging the relative displacements between two horizontal lines spanning the width of the specimen, placed symmetrically with respect to the crack plane. The distance between the lines was $1.2\cdot D$ so that it scaled with the specimen size. Thanks to the use of DIC, the effects of the compliance of the machine were removed. 

\begin{figure}
 \centering
  \includegraphics[trim=0cm 0cm 0cm 0cm, clip=true,width = 0.8\textwidth]{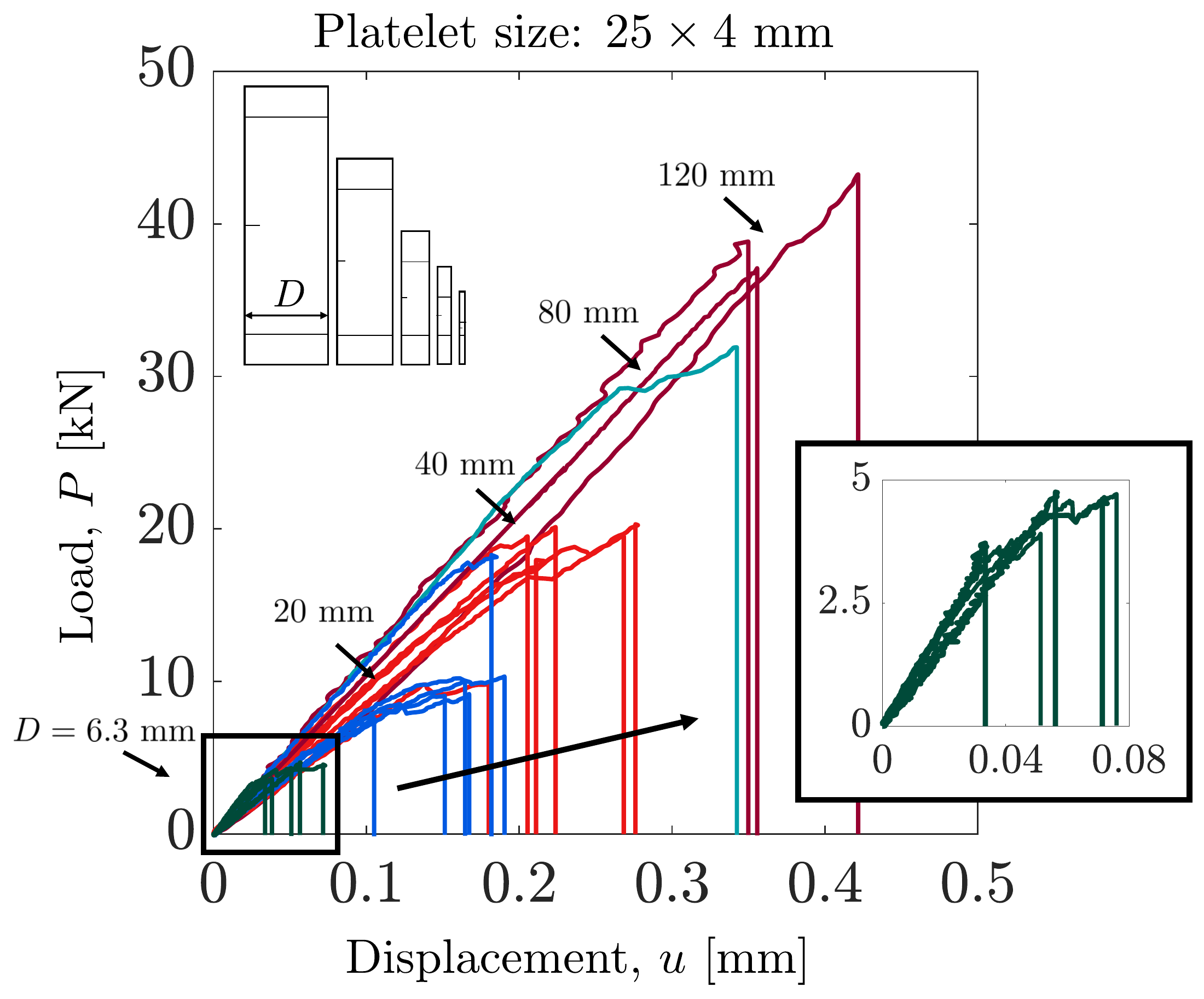} \caption{\label{f3} \sf Representative load-displacement curves of the DFC specimens with the platelet size of $25\times4$ mm.}
\end{figure}

Typical load-displacement curves for the various specimen sizes are shown in Fig.~\ref{f3} for the $25\times 4$ mm platelet. As expected, the initial stiffness is similar for all the specimens. However, in agreement with previous results on quasibrittle materials \cite{Baz1984, Baz1990, Baz1996, Baz1998, Sal2016a, Yao2018a, Yao2018b, Cusatis2018}, a strong size effect can be observed qualitatively by analyzing the structural behavior. That is, for the largest specimen (size-1), the load-displacement curves feature a significantly linear behavior up to the peak load after which sudden failure occurs followed by snap-back instability \cite{Baz1991}. This behavior, representing the typical response of a brittle structure, is in sharp contrast with the smaller sizes. In fact, as the specimen size decreases, the structural response becomes nonlinear before the peak due to sub-critical damage in the FPZ. As can be noted from the inset of Fig.~\ref{f3}, the smallest specimens (size-5) indeed exhibit a remarkable nonlinear response before the peak, this trend being present for all the different platelet sizes investigated.

\subsection{Fracture surfaces}
\begin{figure}
 \centering
  \includegraphics[trim=0cm 0cm 0cm 0cm, clip=true,width = 1\textwidth, scale=1]{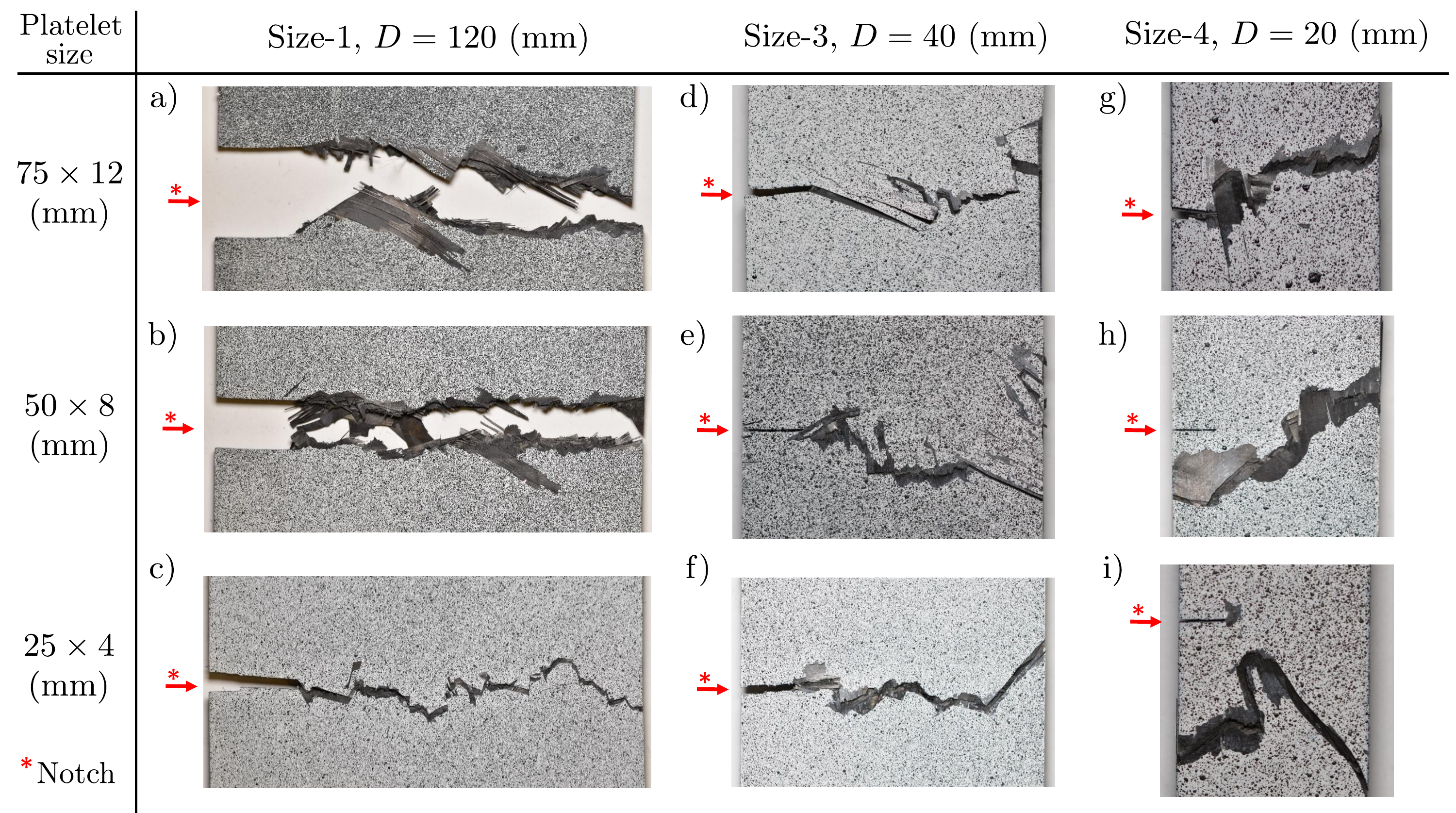} \caption{\label{f4} \sf Representative fracture surfaces of the Single Edge Notch Tension DFC specimens. The red arrow indicates the initial location of the notch.} 
\end{figure}

In Fig.~\ref{f4}, representative fracture surfaces of the tested DFCs are shown. Because of the randomly oriented platelets, mixed damage mechanisms were observed including delaminations between platelets, fiber breakages, pull-outs, and splittings. The fracture paths were also distinct. For size-1, the fracture paths were perpendicular to the loading direction (see Figs.~\ref{f4}a-c) whereas the fracture paths became more chaotic and torturous (see Figs.~\ref{f4}g-i) as the specimen sizes decreased.
For size-4 and -5, several failures were initiated far from the notch, an indication of pronounced quasi-ductile behavior. Indeed, this phenomenon is related to the distributed damage in the FPZ which promotes stress redistribution in front of the notch. In this context, the fracture process may be triggered by other weak spots in the specimen such as resin rich areas, air pockets, or spots with poorly oriented platelets in which the fiber orientations were mainly towards transverse direction with respect to the loading direction.

In previous studies by Feraboli \emph{et al.} \cite{Fera2009c} and Qian \emph{et al.} \cite{Qian2011}, it was claimed that DFC structures can fracture away from a center notch or hole depending on the ratio between the notch and the width of the specimen. However, this statement is only partially true. 
Indeed, the failure location and fracturing behavior depend strongly on the structure geometry (e.g. ratio between notch size and structure width, the shape of the notch etc). Yet, another important aspect of driving the failure behavior is the structure size relative to the platelet dimensions. This is clearly proven in this work since the specimens were all geometrically-scaled and only the characteristic size of the structure was changed. In the experiments, the failure was triggered away from the notch only for size-3, -4, and -5. In such cases, the failure behavior was pseudo-ductile with significant nonlinear energy dissipation before the peak. For all the other larger specimens, the crack always initiated from the notch and the behavior became increasingly more brittle with increasing dimensions.

\subsection{Size effect on the structural strength}
\begin{figure}
 \centering
  \includegraphics[trim=0cm 0cm 0cm 0cm, clip=true, width = 1\textwidth, scale=1]{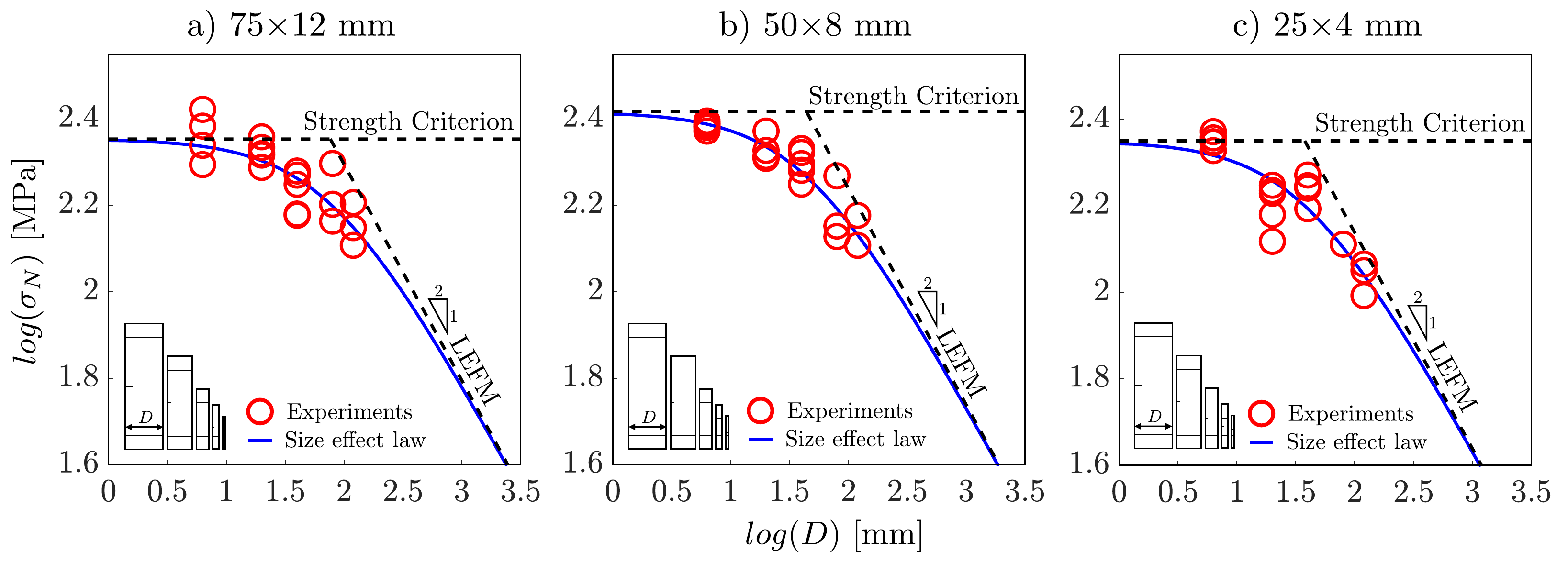}
  \caption{\label{f5} \sf Nominal strength against structure size in double-logarithmic scale showing size effect in DFCs with platelet size of (a) $75\times 12$ mm, (b) $50\times 8$ mm, and (c) $25\times 4$ mm.}
\end{figure}

From the peak load $P_c$ measured during the tests, a nominal strength $\sigma_{Nc}=P_c/tD$ where $t=$ thickness and $D=$ specimen width was calculated. The average nominal strength for various specimens and platelet sizes tested is listed in Table~\ref{T2} and plotted in Figs.~\ref{f5}a-c. Figs.~\ref{f5}a-c show the experimental strength $\sigma_{Nc}$ against the size $D$ in double-logarithmic scale. As can be noted, the figures contain two asymptotes. The horizontal asymptote represents the nominal strength as predicted by stress-based failure criterion, whereas the oblique asymptote with a slope of $-1/2$ represents the prediction by LEFM. It can be noted that, regardless of the platelet sizes, the strength of the SENT specimens decreases as the structure size increases. As mentioned before, this size effect cannot be captured by strength-based failure criteria such as Maximum stress, Tsai-Wu, or others since they predict a constant nominal strength for geometrically-scaled structures. The decreasing strength of experimental data in Figs.~\ref{f5}a-c certainly deviates from the horizontal asymptote representing the strength-based failure criteria. On the other hand, the trend neither follows LEFM, which predicts a scaling of nominal strength with $D^{-1/2}$. To capture the transition from quasi-ductile fracture (displayed by smaller specimens) to brittle fracture (displayed by larger specimens), a theory equipped with a characteristic length-scale related to the size of the FPZ is needed. Such a theory, based on a combination of equivalent fracture mechanics and stochastic finite element modeling, is presented in the following sections.



\section{Analysis and Discussion}

In this study, the fracturing behavior of DFC structures is analyzed using a fracture mechanics approach which implements a characteristic length-scale to capture the effects of the FPZ.

\subsection{Analysis of fracture tests by Size Effect Law (SEL)}

With reference to Fig.~\ref{f2}, let $a_0$ be the initial notch length.
To account for the effects of the nonlinear damage in the FPZ, an equivalent crack of length:
\begin{equation}\label{eq1}
a = a_0 + c_f
\end{equation}
is considered, $c_f$ being an effective FPZ length treated as a material property. 

Following LEFM, the energy release rate $G(\alpha)$ is a function of the crack length as follows:
\begin{equation}\label{eq2}
G(\alpha) = \frac{\sigma_N^2 D}{E^*}g(\alpha)
\end{equation}
with $\alpha = a/D = $ dimensionless crack length, $\sigma_N=$ the nominal stress defined as $\sigma_N = P/(tD)$, and $E^* = $ effective elastic constant. The function $g(\alpha)$ represents the dimensionless energy release rate which relates the geometric and elastic parameters of the structure to $G$ \cite{Baz1998,Sal2016a}. 

In the condition of incipient fracture, the energy release rate $G$ must be equal to the fracture energy $G_f$, assumed to be a material property. By substituting Eq.~(\ref{eq1}) in Eq.~(\ref{eq2}), $G_f$ can be expressed in terms of the effective crack length as follows:
\begin{equation}\label{eq3}
G_f = G(\alpha_0 + c_f/D) = \frac{\sigma_{Nc}^2 D}{E^*}g(\alpha_0 + c_f/D)
\end{equation}

For homogeneous structures, $g=g(\alpha)$ or, in other words, the dimensionless functions depend only on the geometry of the structure \cite[e.g]{Baz1998, Baz1996}. This means that the dimensionless functions take the same value for all the geometrically-scaled structures investigated in the size effect tests. However, this is not the case for DFCs which are highly inhomogeneous. Indeed, the inhomogeneity size is often comparable to the structure size. In this case, the dimensionless energy release functions $g$ and $g'$ may depend on the structure dimension $D$ relative to platelet size and structure thickness. This is because these functions are related to the amount of strain energy stored in the material that is released by the creation of new crack surface area. This quantity is not only influenced by the structure geometry but also by the morphology of the platelets and their orthotropic elastic properties. Since the platelet size is not scaled with the structure and the number of platelets may not be enough to make the structure statistically homogeneous, a size effect on the energy release occurs. 

In view of these considerations, Eq.~(\ref{eq3}) must be rewritten as follows:
\begin{equation}\label{eq4}
G_f = \frac{\sigma_{Nc}^2 D}{E^*}g(\alpha_0 + c_f/D, D)
\end{equation}
where now the dimensionless energy release is considered as a function of both the relative crack length and structure size: $g=g\left(\alpha,D\right)$. Performing a Taylor series expansion around $\alpha_0$ for a constant structure size $D$ one gets:

\begin{equation}\label{eq5}
G_f = \frac{\sigma_{Nc}^2 D}{E^*}\Big[g(\alpha_0, D) + \frac{c_f}{D}\frac{\partial g}{\partial \alpha}(\alpha_0, D)\Big]
\end{equation}

After rearranging Eq.~(\ref{eq5}), the following equation, known as \B's Size Effect Law (SEL) \cite{Baz1984, Baz1996, Baz1998} modified for DFC structures is obtained:

\begin{equation}\label{eq6}
\sigma_{Nc} = \sqrt[]{\frac{E^* G_f}{Dg(\alpha_0, D)+c_f g'_D(\alpha_0, D)}}
\end{equation}
where $g'_D=\left[\partial g/\partial \alpha\right]_D$ and the subscript $D$ indicates partial differentiation with a constant structure size. As can be noted, the main difference compared to the homogeneous case, Eq.~(\ref{eq3}), is that the new equation features an additional size effect related to the dimensionless energy release functions. These functions can be calculated leveraging the stochastic finite element framework presented in the following sections.

Different from LEFM, the foregoing equation relates the nominal strength not only to the fracture energy of the material but also a characteristic length scale $c_f$, associated to the FPZ size. This length scale is the key to capture the transition of the fracturing behavior from quasi-ductile to brittle with increasing structure size. Finally, the previous expression can be also written as follows:
\begin{equation}\label{eq7}
\sigma_{Nc} = \frac{\sigma_0}{\sqrt[]{1+D/D_0}}
\end{equation}
where $\sigma_0 = \sqrt{E^*G_f/c_fg'(\alpha_0, D)}$ and $D_0 = c_fg'(\alpha_0, D)/g(\alpha_0, D)$. $\sigma_0$ and $D_0$ are size effect constants depending on the structure geometry and the size of the FPZ.

It is worth noting that for the structure sizes and the thickness considered in this study, the average dimensionless energy release functions were not found to change significantly with $D$ based on an extensive computational study (detailed discussion in section 5.2). Therefore, for the sake of simplicity, Eq.~(\ref{eq6}) was used with the average value of $g(\alpha_0)$ and $g'(\alpha_0)$ of all the specimen sizes for a given platelet size. However, the reader should keep in mind that for all the cases in which $g$ and $g'$ are shown to depend significantly on the structure dimensions $D$, Eq.~(\ref{eq6}) with $g(\alpha_0, D)$ and $g'(\alpha_0, D)$ should be used. 

Finally, it is worth concluding this section stressing the simplicity of Eq.~(\ref{eq6}). This expression does account for all the key aspects of the fracturing process in DFCs. While the effect of the nonlinear cohesive stresses in the FPZ is accounted for by the introduction of the length scale $c_f$, the effects of the platelet size and morphology on the energy release are captured by the functions $g$ and $g'_D$. These functions are calculated by explicitly modeling the platelets by FEA and characterizing the evolution of the strain energy in the structure for different crack lengths.

\subsection{Fitting of the experimental data using the SEL}
\begin{figure}
 \centering
  \includegraphics[trim=0cm 0cm 0cm 0cm, clip=true,width = 1\textwidth, scale=1]{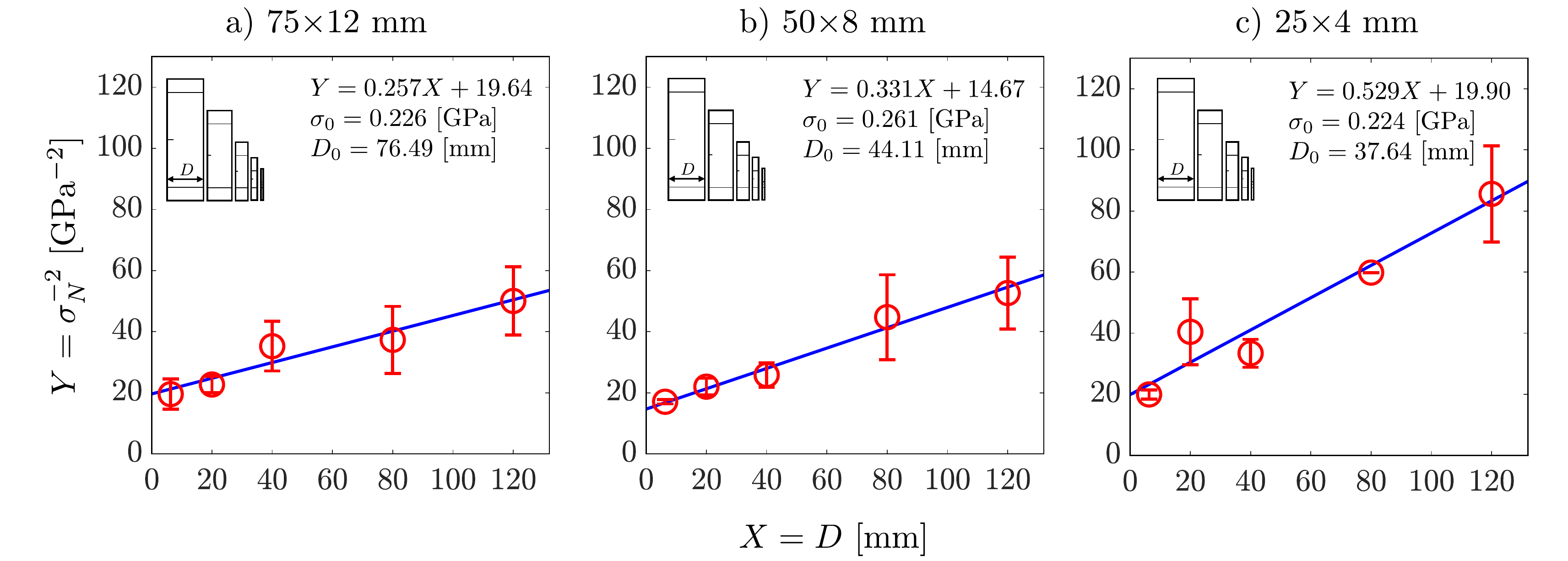} \caption{\label{f6} \sf Linear regression analysis to find the size effect law parameters, $\sigma_0$ and $D_0$ for the platelet size of a) $75\times 12$ mm, b) $50\times 8$ mm, and c) $25\times 4$ mm.} 
\end{figure}

To obtain the size effect parameters $\sigma_0$ and $D_0$, regression analysis was conducted on the experimental data. To do so, the following transformation was used:
\begin{equation}\label{eq8}
X = D, \;\; Y = \sigma_{Nc}^{-2}
\end{equation}
Using these terms, Eq.~(\ref{eq6}) can be expressed in the following linear form:
\begin{equation}\label{eq9}
Y = C + AX
\end{equation}
with:
\begin{equation}\label{eq10}
\sigma_0 = C^{-1/2}, \;\; D_0 = \frac{C}{A} = \frac{1}{A(\sigma_0)^2}
\end{equation}
Leveraging this equation, it is possible to perform a linear regression of the size effect data as shown in Figs.~\ref{f6}a-c for all the investigated platelet sizes. Then, the size effect parameters, $\sigma_0$ and $D_0$ are extrapolated from the y-intercept and the slope of the linear regression, by means of Eqs.~(\ref{eq10}).



Figs.~\ref{f5}a-c show the fitting into SEL based on Eq.~(\ref{eq6}). As can be noted, the results for all the platelet sizes are characterized by a significant deviation from LEFM, the deviation being more pronounced for smaller specimens and larger platelets. In particular, the figures show a transition of the experimental data from stress-driven failure, characterized by the horizontal asymptote, to energy-driven fracture characterized by the $-1/2$ asymptote. This phenomenon can be ascribed to the increased size of the FPZ compared to the structure size, which makes the non-linear effects caused by micro-damage in front of the crack tip not negligible. For sufficiently small specimens, the FPZ affects the structural behavior and causes a significant deviation from the scaling effect predicted by LEFM. Indeed, the structural strength is less affected by the size. On the other hand, for increasing sizes, the effects of the FPZ become less significant, thus leading to a stronger size effect closely captured by LEFM. Further, comparing the size effect plots of DFCs from small to large platelet sizes, a gradual shift of experimental data points can be noted towards the quasi-ductile region. Thus, it shows that not only the larger platelets lead to a higher fracture toughness but also to a gradually more quasi-ductile structural behavior for a given size. 

The foregoing conclusions are extremely important for the design of DFC structures featuring defects or sharp notches. The pseudo-ductile behavior reported in fracture tests on small laboratory-scaled DFC specimens may induce designers to overestimate severely the load capacity of real, large structural components if strength-criteria is used. On the other hand, LEFM does not always provide an accurate method to extrapolate the structural strength of larger structures from lab tests on small-scale specimens, especially if the size of the specimens belongs to the transitional zone. In fact, the use of LEFM in such cases may lead to a significant underestimation of structural strength, thus hindering the full exploitation of DFC fracture properties. This is a severe limitation in several engineering applications such as aeronautics and astronautics for which structural performance optimization is of utmost importance. On the other hand, LEFM always overestimates significantly the strength when used to predict the structural performance at smaller length-scales. This is a serious issue for the design of e.g. small complex-shaped components. In such cases, SEL or other equivalent material models with a characteristic length scale ought to be used.

\begin{figure}
 \centering
  \includegraphics[trim=0cm 0cm 0cm 0cm, clip=true,width = 1\textwidth, scale=1]{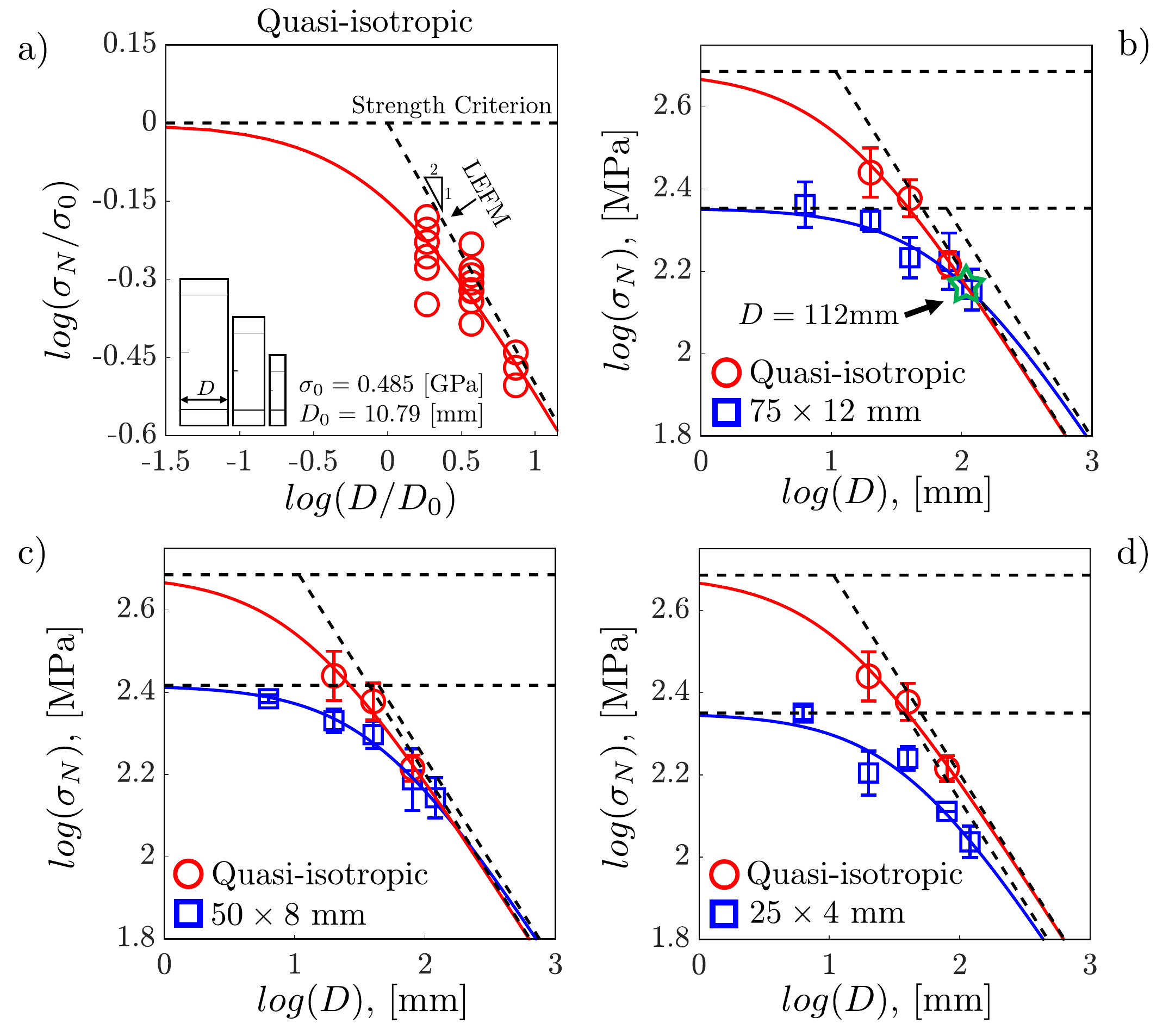} \caption{\label{f7} \sf a) Normalized size effect curve of quasi-isotropic (QI) laminate, b) size effect curve of QI with $75\times15$ mm, c) QI with $50\times8$ mm, and d) QI with $25\times4$ mm.} 
\end{figure}

It is interesting to compare the structural behavior of DFCs to traditional continuous composite structures. To do so, quasi-isotropic (QI) laminates $[0/45/90/-45]_{3s}$ with geometrically-scaled SENT specimens were manufactured with the identical prepreg system. The average thickness of the QI laminates were $3.43$ mm. The test method was identical with the DFCs. The resulting strengths are listed in Table~\ref{T2}. Using the Eq.~(\ref{eq6}), the size effect curve of the QI laminate is plotted in Fig.~\ref{f7}a. The QI possesses far more brittle behaviors compared to the DFCs reaching oblique asymptote even with the smallest specimen size. Because of the higher sensitiveness to the stress risers in the QI laminate, the strength drops much quickly as the structure size increases. Surprisingly, the QI provides lower strength after reaching the structure size of $112$ mm comparing with the DFCs made of largest platelet size (see the star mark in Fig.~\ref{f7}b). This result indicates that DFCs are more suitable than QI laminates for the structures with sufficiently large size containing the geometrical stress risers such as notches or holes. 

\begin{figure}
 \centering
  \includegraphics[trim=0cm 0cm 0cm 0cm, clip=true,width = 0.8\textwidth, scale=1]{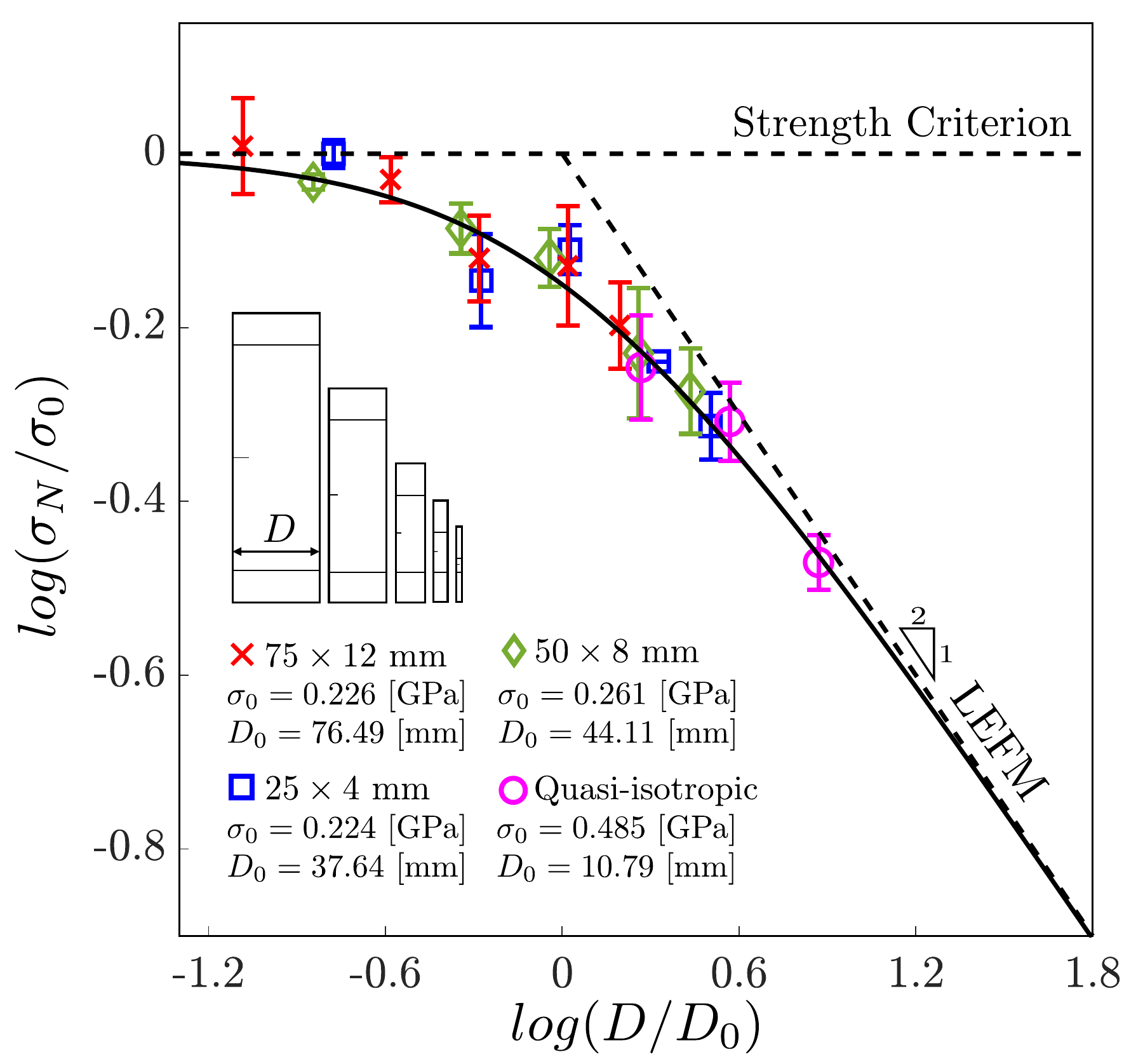} \caption{\label{f8} \sf Normalized size effect curves in double-logarithmic scale showing a size effect in the DFCs with all the platelet sizes and the quasi-isotropic laminate.}
\end{figure}

Figure~\ref{f8} combines all the normalized experimental strength, $\sigma_{Nc}/\sigma_0$, against the normalized size ,$D/D_0$, in double-logarithmic scale with the SEL fitting. Regardless of the platelet sizes, all the normalized strengths are well captured by the SEL. Also, the structure sizes tested in the laboratory are within the neighbor of $D_0$, where the two asymptotes meet. This $D_0$ is called the \textit{transitional size} where it locates the transition from quasi-ductile to brittle behavior. This trend confirms that DFCs are the quasibrittle material containing the non-negligible size of the FPZ.

\subsection{Brittleness number of DFCs vs traditional laminated composites}
\begin{figure}
 \centering
  \includegraphics[trim=0cm 0cm 0cm 0cm, clip=true,width = 1\textwidth, scale=1]{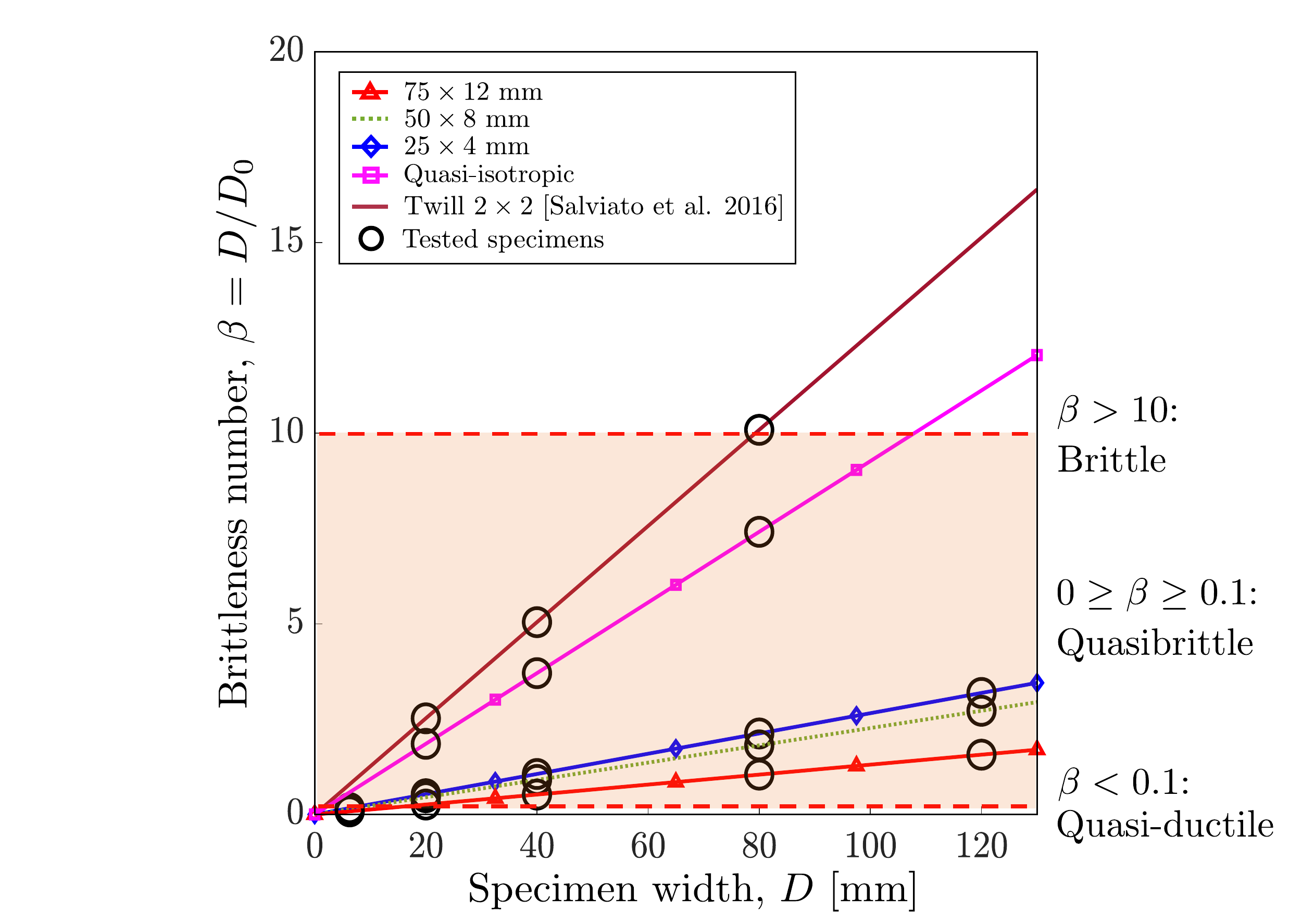} \caption{\label{f9} \sf Brittleness number, $\beta$, vs structure size for the DFCs investigated in this work, a quasi-isotropic laminated composite made using the same prepregs, and a carbon twill $2\times 2$ composite \cite{Sal2016a}.}
\end{figure}

For additional structural behavior comparison between the traditional composite structures with DFCs, a useful non-dimensional parameter called the \textit{brittleness number}, $\beta$, is introduced \cite{Baz1998}. This parameter, comparing the brittleness of structures with similar geometry and size, is defined as the ratio between the structure characteristic size, $D$, and the transition size, $D_0$. When $\beta$ is greater than $\sim 10$, the behavior of the structure is typically very brittle and LEFM is well suitable to capture the fracturing behavior of the material. When $\beta$ is less than $0.1$, the structure can be considered as quasi-ductile or perfectly plastic. The stress-based failure criteria provides a fairly good prediction of the structural strength. If $\beta$ lies in between $10$ and $0.1$, the structure should be treated as quasibrittle material. Figure~\ref{f9} shows $\beta$ for the DFC structures investigated in this work compared to a QI laminate. Further, the $\beta$ for the textile composite structures tested in \cite{Sal2016a} is also provided for reference. As can be noted, the brittleness number of DFCs is always within the boundary of the quasibrittle zone for all the structure sizes investigated, regardless of the platelet size. In contrast, both the QI and textile composite feature higher brittleness numbers for the same structure size, with the largest specimen reaching the LEFM region. 
From the foregoing analysis, it is evident that DFCs are, by far, less brittle than traditional composites. This is a highly desirable condition for the design of damage tolerant composite structures featuring notches and other stress raisers. 


\section{Stochastic finite element model}
One of the main objectives of this study is to estimate, for the first time, the mode I intra-laminar fracture energy $G_f$ of DFCs. To characterize $G_f$ leveraging Eq.~(\ref{eq6}), the dimensionless functions $g(\alpha_0)$ and $g'(\alpha_0)$ need to be calculated. These functions are related to the release of elastic strain energy induced by the crack, which is significantly influenced by the platelet constitutive properties and, more importantly, the random platelet distribution. To capture these aspects, a stochastic finite element model is proposed in the following sections.

\subsection{Mesostructure generation}\label{algo}
The mesostructure generation algorithm adopted in this work represents an extension of the stochastic laminate analogy method proposed in \cite{Fera2010, Tuttle2017, Sel2017}. 

\begin{figure}
 \centering
  \includegraphics[trim=0cm 0cm 0cm 0cm, clip=true,width = 0.8\textwidth, scale=1]{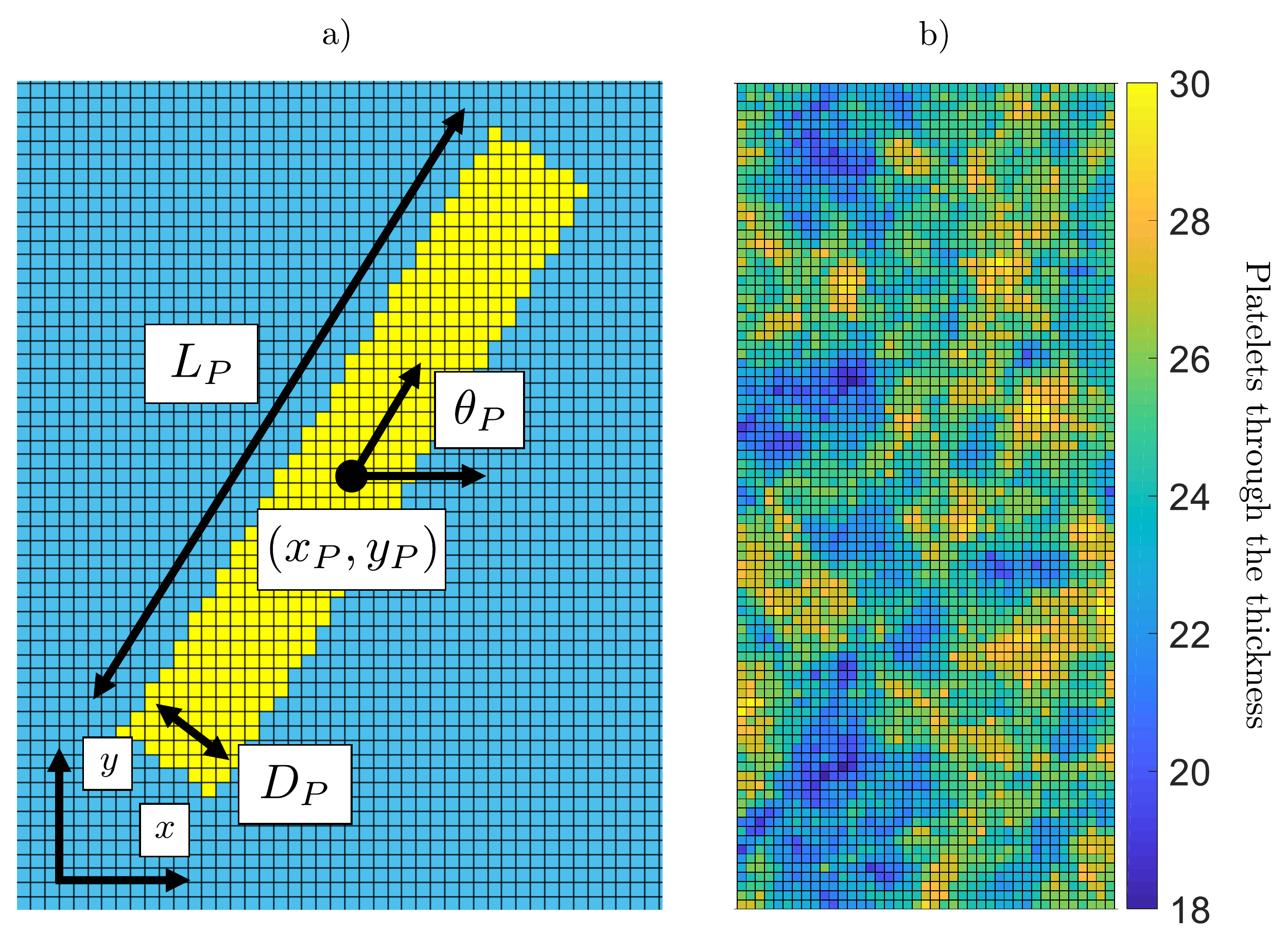} \caption{\label{f10} \sf a) A detailed platelet geometry. Here, $L_P$ and $D_P$ represent the length and width of the platelet, respectively. The platelet distribution algorithm generates the coordinate ($x_P$, $y_P$) and orientation angle ($\theta_P$) of a platelet via uniform probability distribution. b) sample DFC mesostructure with the average platelets through the thickness = $24$. }
\end{figure}

Following the foregoing computational framework, the structure is first subdivided into several partitions, whose size is chosen to allow an accurate description of the platelets. In the present work, this size was set to be $1\times1$ mm for all the platelet sizes investigated \cite{Sel2017}. Once the platelet length and width are chosen, the algorithm generates the platelets over the structure (see Fig.~\ref{f10}a). A probability distribution can be used to assign both the center points and the orientations of the platelets. A thorough morphological study confirmed that a uniform probability distribution provided the best compromise between the accurate description of the mesostructure and the simplicity of the theoretical framework for the specimens investigated in this work. It is worth mentioning, however, that such a probability distribution can be far from uniform in case of significant platelet flow and complex structures. In such cases, the probability distribution should be characterized experimentally or estimated by simulating the manufacturing process numerically \cite{Denos2018a, Denos2018b, Sommer2017}. 

The algorithm repeats until the average number of platelets through the thickness reaches the desired value (see Fig.~\ref{f10}b). In this study, $3.3$ mm corresponds to $24$ layers. However, there are two main issues that the platelet generation needs to clarify. First, without any restriction, the spatial variability of the number of platelets through the thickness would be unrealistically extreme. Second, since the resin flow during the manufacturing process is not modeled explicitly, the algorithm would result in an uneven thickness of the plate unless the platelets and added resin rich layers are scaled properly. To address these problems, an in-house DFC mesostructure generation algorithm is developed, as outlined in Fig.~\ref{f11}. Details of this algorithm are explained in the following sections.

\subsubsection{Platelet distribution algorithm}
\begin{figure}
 \centering
  \includegraphics[trim=0cm 0cm 0cm 0cm, clip=true,width = 1\textwidth, scale=1]{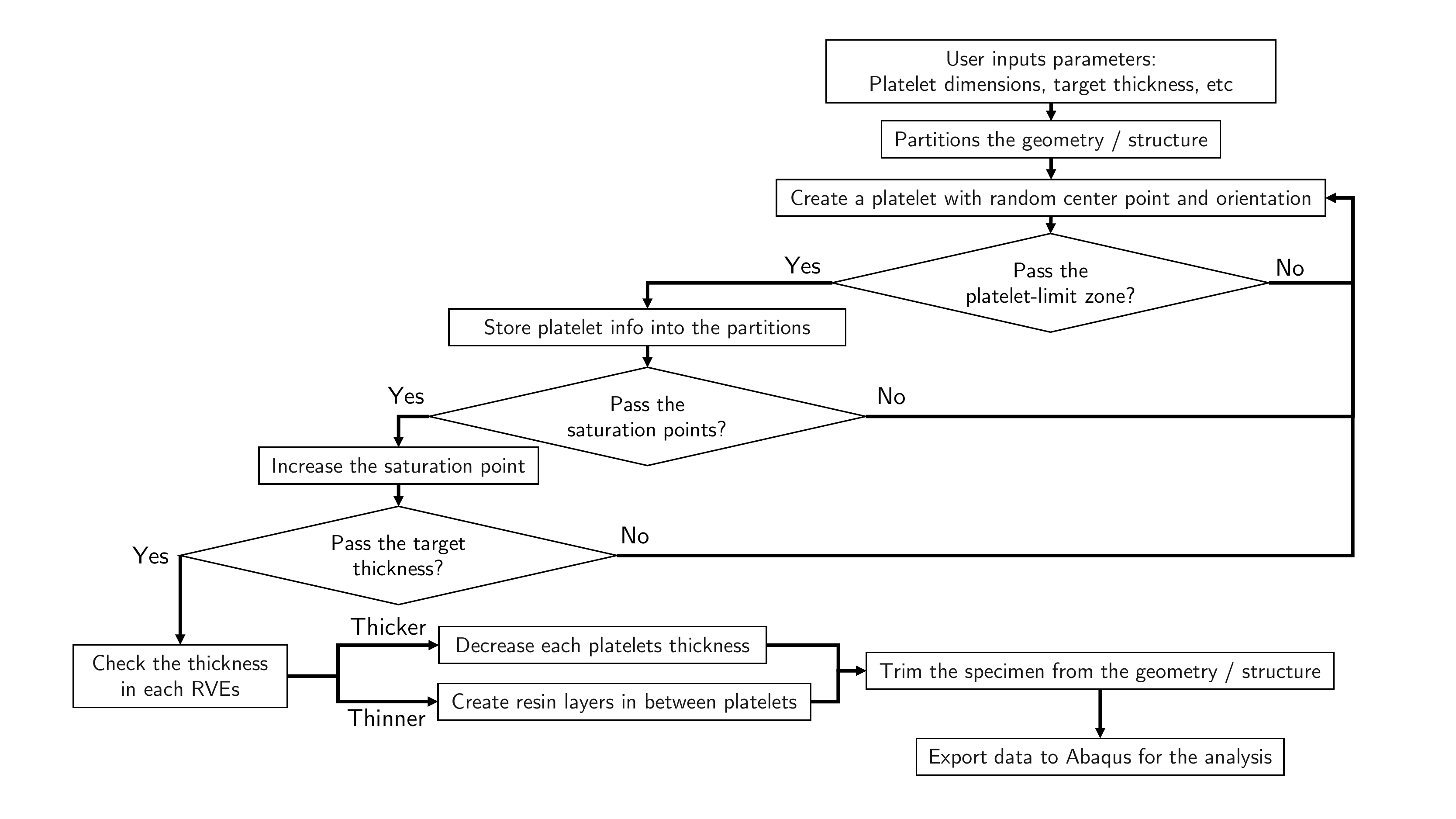} \caption{\label{f11} \sf The flow-chart of the mesostructure generation algorithm.}
\end{figure}

DFC structures exhibit a natural spatial distribution on the number of platelets through the thickness. To characterize this morphological feature, $90$ DFC samples were observed under an optical microscope (Omax 9 MP camera with Nikon M Plan 20 objective) to record the actual number of platelets through the thickness. The measured coefficient of variation (CoV) was $0.22$ for an average of $24$ layers. To achieve the desired CoV in the numerical model, two strategies are implemented in the algorithm: (1) the platelet-limit zone and (2) the saturation points (see the decision boxes in Fig.~\ref{f11}). The platelet-limit zone blocks additional platelet depositions at the certain partitions when they reach the allowable limit. If a platelet is placed on top of the limit zone, the algorithm rejects it and a new platelet is generated at another random location. The saturation points decide the allowable limits in the partitions. In the case of $24$ layers, every third layer is the saturation point which corresponds to the mat stacking manufacturing method as described in Section 2.1. Therefore, the allowable limits are $[3, 6, 9, ... 24]\times$CoV. The algorithm repeats the platelet generations until the number of platelets through the thickness reaches the saturation points. If the condition is met, the saturation point moves to the next point until the final limit point ($=24$). 

\begin{figure}
 \centering
  \includegraphics[trim=0cm 0cm 0cm 0cm, clip=true,width = 1\textwidth, scale=1]{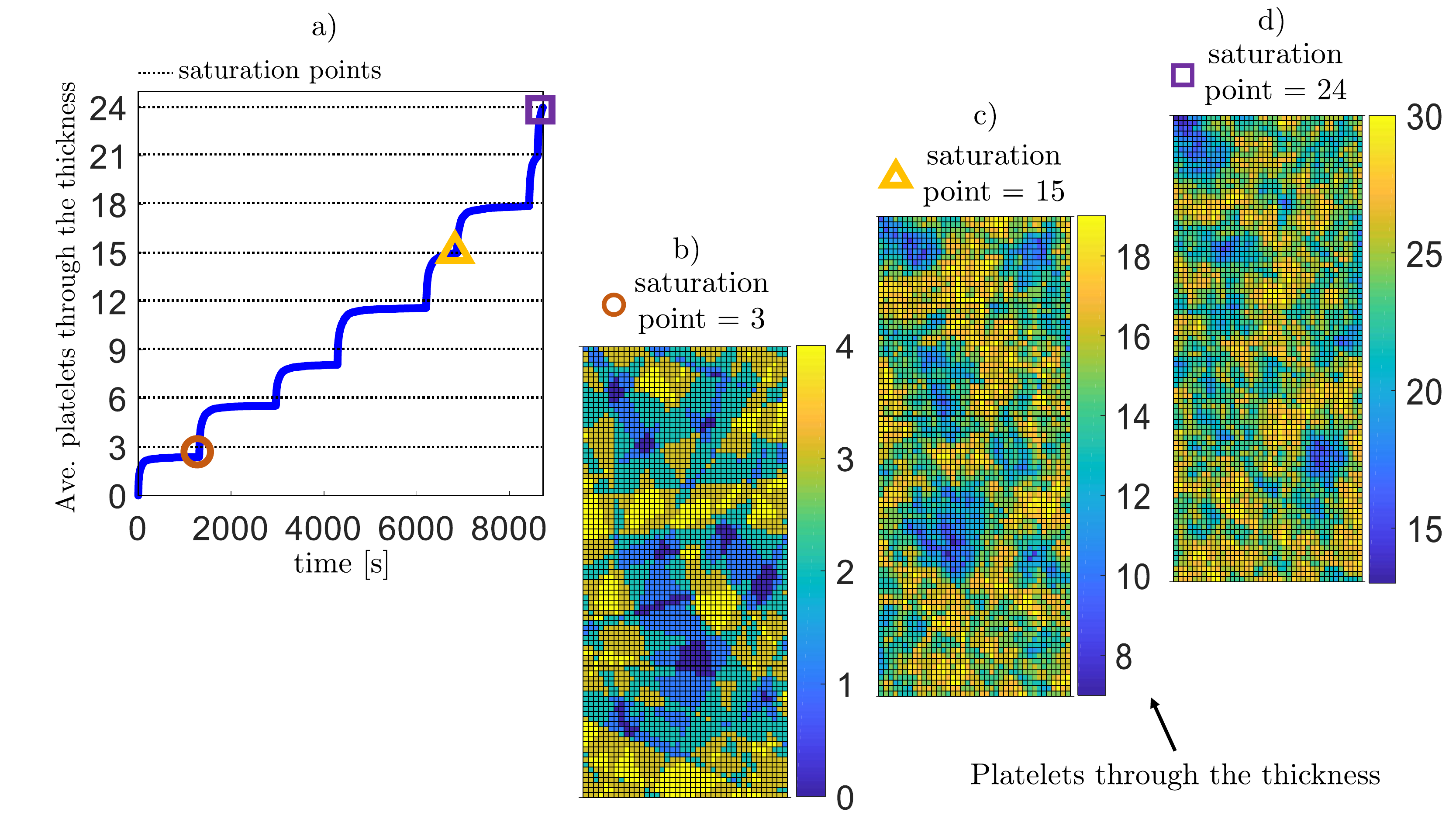} \caption{\label{f12} \sf The platelet distribution scheme. a) The proposed saturation point method, and mesostructure at the saturation point of b) $3$, c) $15$, and d) $24$.}
\end{figure}

Figure~\ref{f12}a shows the evolution of the average platelets as a function of the running time of the platelet distribution algorithm. A cascading trend is observed, with each step corresponds to the saturation points. The advantage of this algorithm is the accurate control of the CoV, resulting in a more realistic mesostructure as shown in Figs.~\ref{f12}b-d. As the result, the maximum and minimum numbers of the platelets through the thickness are 30 and 18 or 30 and 15 at the end of the platelet generation (see the color bar in Figs.~\ref{f10}b or~\ref{f12}d). Without such platelet distribution tactics, this contrast of platelet numbers can be far off from measured CoV, which is unrealistic. 

\subsubsection{Platelet thickness adjustment algorithm}
\begin{figure}
 \centering
  \includegraphics[trim=0cm 1cm 0cm 0cm, clip=true,width = 1\textwidth, scale=1]{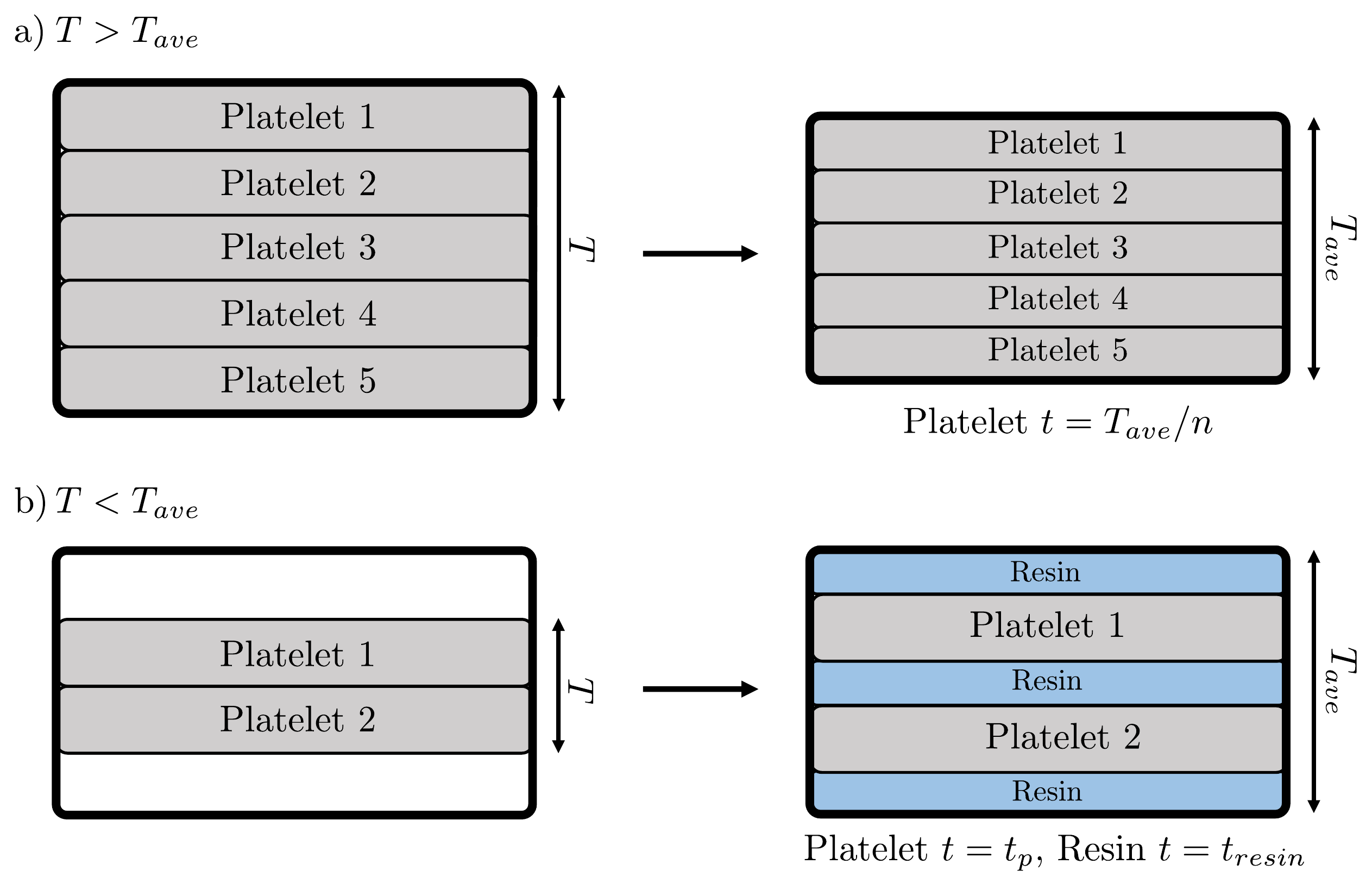} \caption{\label{f13} \sf The platelet thickness adjustment scheme a) when the platelet thickness was reduced, b) when the resin layers were introduced.}
\end{figure}

Once the mesostructure is constructed within the desired CoV, the algorithm proceeds with the adjustment of the partitions thickness. This step is made necessary by the fact that the resin flow is not modeled explicitly. Accordingly, at the end of the algorithm, each partition will be characterized by a different thickness, in contrast to the fact that real DFC plates feature an almost uniform thickness. To overcome this issue, two possible conditions are addressed. (1) when the number of platelets exceeds the target average number, each platelet thickness is reduced linearly to match the target thickness (see Fig.~\ref{f13}a). This is assumed to be a valid assumption since, in reality, the platelets are compressed and spread out by the hot press machine to meet the average thickness. The deformation of the platelet geometry is neglected in this study. (2) If the number of platelets is lower than the target, instead of increasing the platelet thickness, layers of resin are introduced (see Fig.~\ref{f13}b). This is done to indirectly mimic the flow of the resin into the lower thickness regions (i.e. regions with the lower number of platelets). The resin layers are assumed to have the matrix system material properties of the T$700$G.

\subsection{Computation of $g(\alpha)$ and $g'(\alpha)$ and the fracture energy}

The mesostructure generated by the algorithm described in section~\ref{algo} was imported in Abaqus/Standard \cite{Abaqus}. $8$-node, quadrilateral Belytschko-Tsay shell elements (S$8$R) were used to model the structure and to calculate the reaction force, $P$, and the total strain energy. The behaviors of the platelets and resin layers were assumed to be linearly elastic, with the elastic properties given in Table~\ref{T3}. 
A uni-axial uniform displacement was applied at one end of the specimen while the other end was fixed in all directions. To find the dimensionless functions $g(\alpha)$ and $g'(\alpha)$ using Eq.~(\ref{eq2}), the energy release rate $G(\alpha)$ needed to be computed using the finite element method. 

\begin{figure}
 \centering
  \includegraphics[trim=0cm 0cm 0cm 0cm, clip=true,width = 1\textwidth, scale=1]{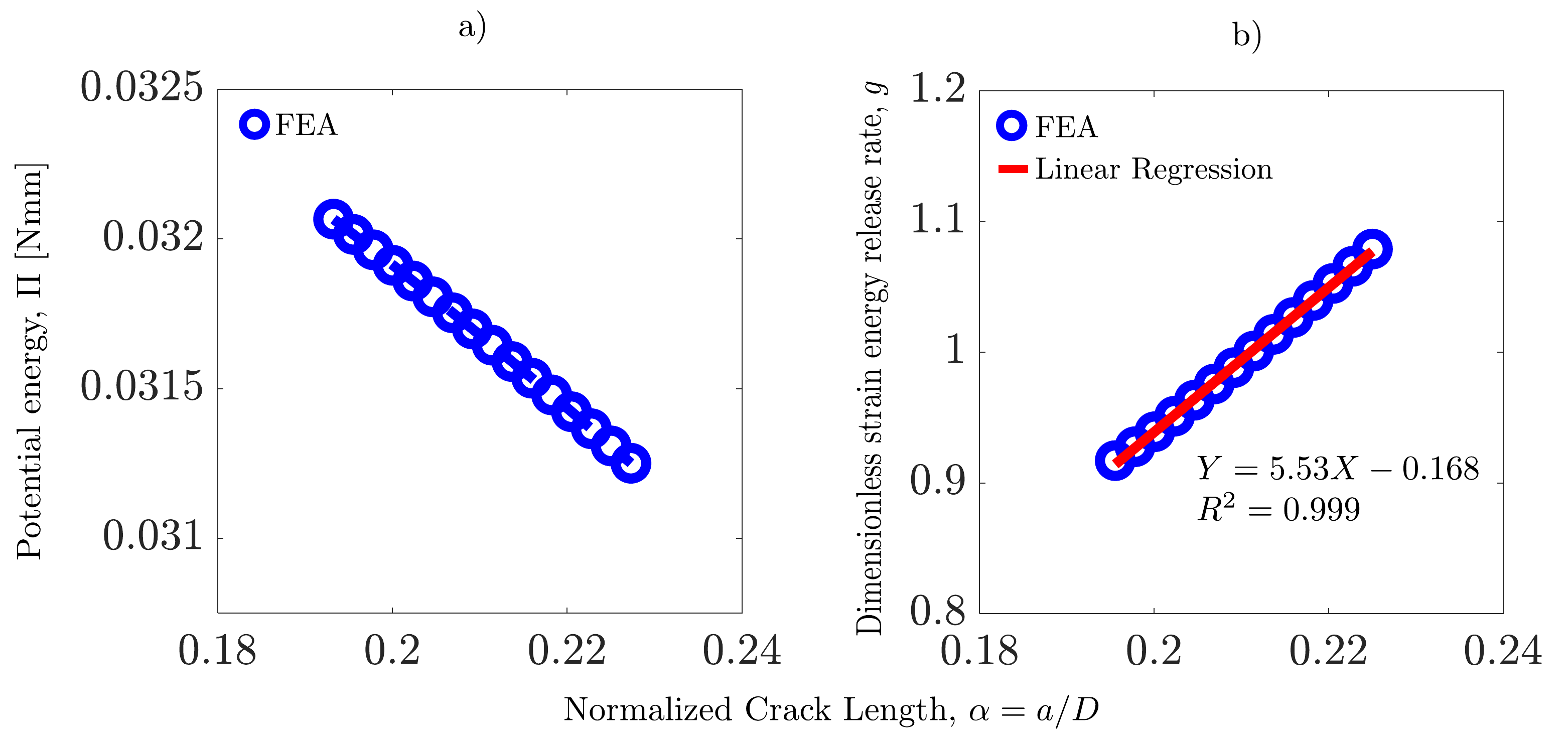} \caption{\label{f14} \sf a) Obtained potential energy of a typical SENT DFC specimen, b) The calculation of dimensionless energy release rate $g(\alpha_0)$ and its derivative $g'(\alpha_0)$.}
\end{figure}

Generally, a convenient method to obtain the energy release rate is using the J-integral approach \cite{Rice1968, Sal2016a}. However, this is not applicable to DFCs due to their in-homogeneous material characteristics. Accordingly, to calculate $G(\alpha)$, its definition is used directly
\cite{Baz1998}:
\begin{equation}\label{eq11}
G(u,a) = -\frac{1}{b}\left[\frac{\partial\Pi(u,a)}{\partial a}\right]_u
\end{equation}
with $u$ being the equilibrium displacement, $a$ being the crack length, $b$ the thickness, and $\Pi$ being the potential energy of the whole specimen. Figure \ref{f14}a shows the potential energy of a typical DFC SENT specimen. Then, $G(u,a)$ is calculated by means of a central finite difference approximation, $G\left(u,a\right)\approx -1/b \left[\Delta \Pi\left(u,a\right)/\Delta a\right]_u$, as a function of the normalized crack length, $\alpha=a/D$.
Once the energy release is calculated, $g(\alpha)$ can be calculated taking advantage of Eq. (\ref{eq2}), whereas $g'(\alpha)$ can be calculated by linear interpolation for $\alpha \rightarrow \alpha_0$ (see Fig.~\ref{f14}b).

\begin{figure}
 \centering
  \includegraphics[trim=0cm 0cm 0cm 0cm, clip=true, width = 1\textwidth, scale=1]{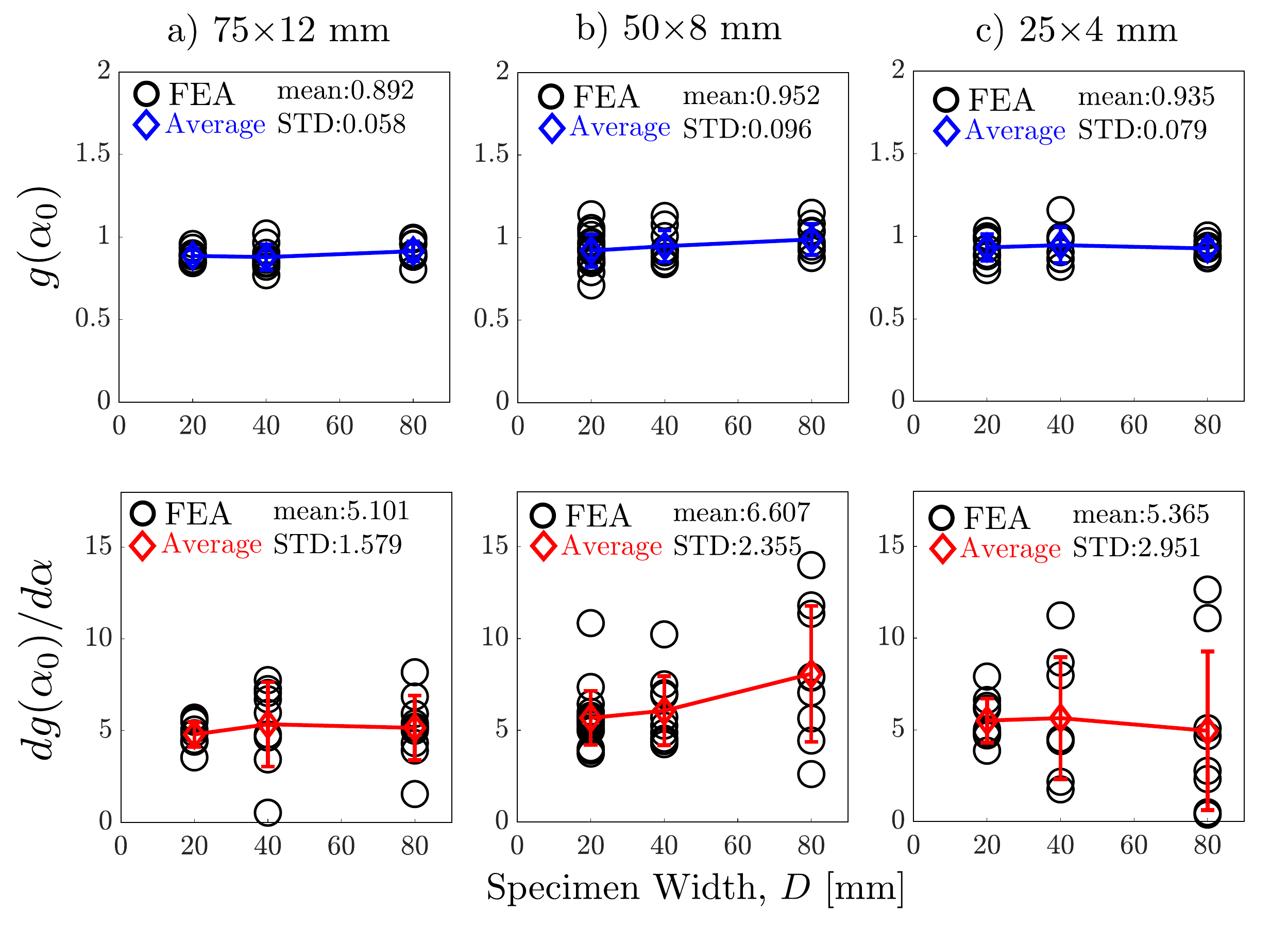}
  \caption{\label{f15} \sf Comparison of dimensionless energy release rate and its derivative for the platelet size of (a) $75\times 12$ mm, (b) $50\times 8$ mm, and (c) $25\times 4$ mm.}
\end{figure}

For homogeneous, geometrically-scaled specimens, the functions $g(\alpha)$ and $g'(\alpha)$ do not depend on the structure size \cite{Baz1998}. However, in the case of DFCs, this is not generally true as explained in the previous sections. For this reason, $10$ different mesostructures of the size-$2$, -$3$, and -$4$ were simulated to verify if any size effect on the dimensionless energy release functions was present for the structures investigated in this work. The results of this computational study are presented in Fig.~\ref{f15} which provides $g(\alpha)$ and $g'(\alpha)$ with $\alpha = \alpha_0$ as a function of the specimen size for the various platelet dimensions. As can be noted, for a moderate window of the specimen and platelet size variations considered in this work, the structural size effect on the average value of these functions is not significant. A slight effect of the size on the CoV of the function $g'$ is noticeable but was neglected in this work as it does not affect the average behavior. Based on these results, the average values of the dimensionless functions were used. These values are summarized in Table \ref{T4} along with the ones for a quasi-isotropic laminate with equal thickness. Though the variations of $g(\alpha, D)$ and $g'(\alpha, D)$ are minute in this study, the case of their large variations will be reported in the future publications.

\begin{figure}
 \centering
  \includegraphics[trim=0cm 0cm 0cm 0cm, clip=true,width = 0.7\textwidth, scale=1]{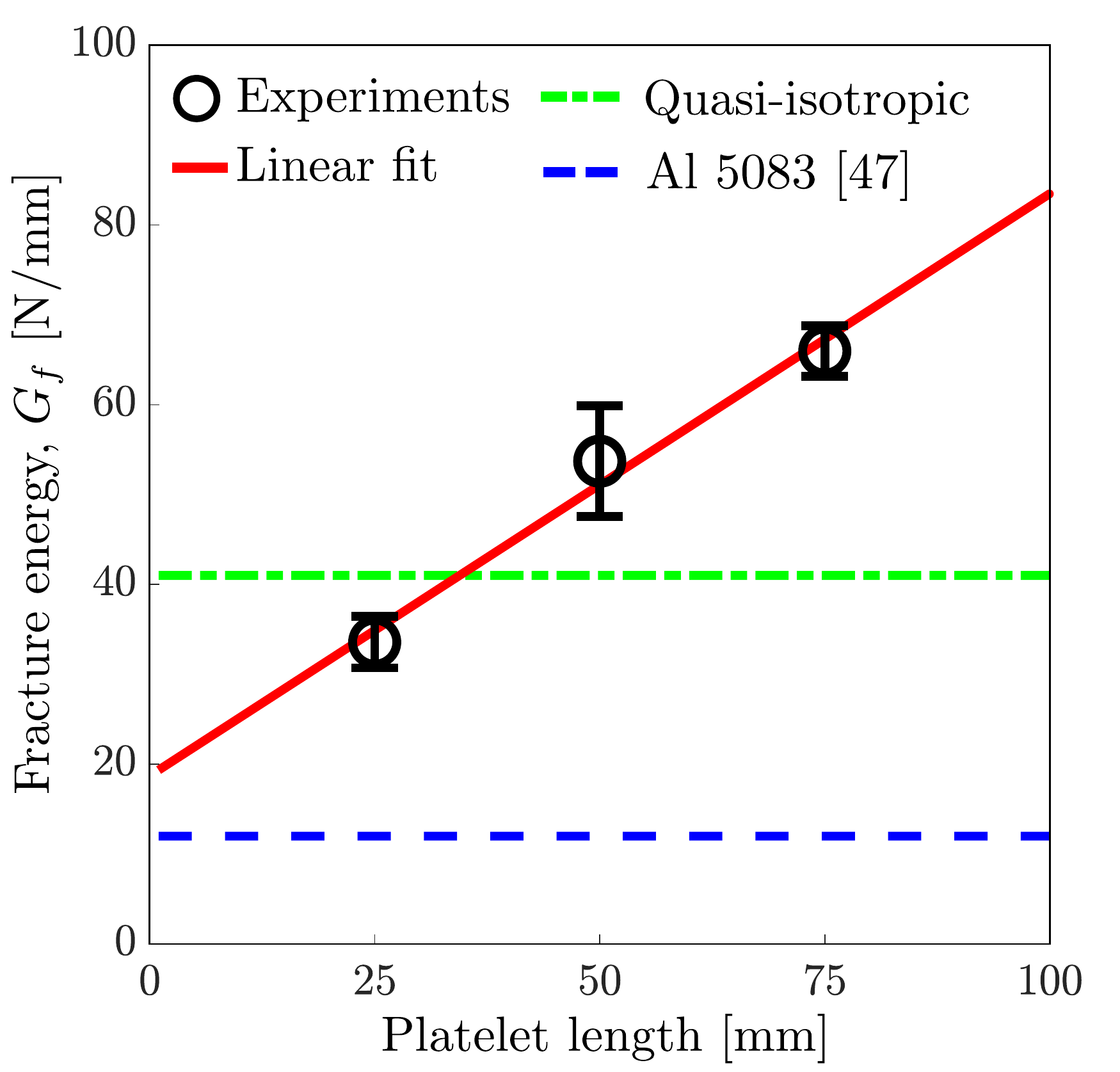} \caption{\label{f16} \sf Comparison of the mode I fracture energy for DFCs, a quasi-isotropic layup, and a typical aluminum alloy.}
\end{figure}

Connecting the size effect parameters, $A$ and $C$, with the dimensionless functions leveraging Eq.~(\ref{eq6}), the fracture energy, $G_f$, and the effective FPZ length, $c_f$, can be calculated:
\begin{equation}\label{eq12}
G_f = \frac{g(\alpha_0)}{E^*A}, \;\;
c_f=\frac{E^*G_fC}{g'(\alpha_0)}
\end{equation}
Table~\ref{T4} lists the fracture parameters for all the platelet sizes. To provide better comparison, the fracture energy of three different platelet sizes is plotted in Fig.~\ref{f16} along with the one of a typical Al5083 \cite{Nikos1987} and of a T$700$G QI laminate composite. The $G_f$ of the QI laminate was obtained experimentally following the identical test procedure.

As can be noted from the Fig.~\ref{f16}, the complex heterogeneous mesostructure of DFCs provides them with the outstanding fracture energy. The fracture energy calculated in this work is $65.47$ N/mm, $55.05$ N/mm, and $33.66$ N/mm for the $75\times12$, $50\times8$, and $25\times4$ mm platelets respectively. These fracture energies are from $5.5$ to $2.8$ times larger than the one of a typical Al5083 ($\sim 12$ N/mm). This is remarkable considering that aluminum is one of the main competing materials for the manufacturing of complex, lightweight aerospace/automobile components. It is also a promising result in view of the possible application of DFCs to improve the crashworthiness of lightweight structures. It is noteworthy that all the DFCs considered in this study exhibit the fracture energy that is comparable or even larger than the QI laminate made from the same materials ($=41.01$ N/mm). For the DFCs made by $75\times12$ mm platelets, the fracture energy is 1.60 times larger.

The results also clearly indicate a strong effect of the platelet size on the fracture properties of the material. In particular, the fracture energy increases linearly with the platelet size in the range investigated in this work. Further computational studies are ongoing to clarify how this trend is affected by the mechanical properties of the constituents and the random orientation of the platelets.
Finally, it is worth mentioning here that the platelet size also affects the effective FPZ length which is found to be $13.21$, $6.41$, and $6.54$ mm for the $75\times12$, $50\times8$, and $25\times4$ mm platelets respectively. This is in agreement with the more pronounced quasi-ductility reported for the DFC structures made of longer platelets.



\section{Conclusion}
This work investigated the fracturing behavior and scaling of Discontinuous Fiber Composite (DFC) structures leveraging a combination of experiments and stochastic computational modeling. Based on the results of this study, the following conclusions can be elaborated:


\be  \setlength{\itemsep}{1.3mm}

\ii The fracture tests on geometrically-scaled Single Edge Notch Tension (SENT) specimens confirmed a remarkable size effect on the structural strength. It was found that, when the specimen is sufficiently large, the structural strength scales according to the Linear Elastic Fracture Mechanics (LEFM) and the failure occurs in a very brittle way. In contrast, small specimens exhibit a more pronounced pseudo-ductility with limited scaling effect and a significant deviation from LEFM.


\ii The transition from pseudo-ductile to brittle fracture with an increasing specimen size is related to the development of a significant Fracture Process Zone (FPZ) whose dimensions were found to be comparable to the platelet size. In the FPZ, signiﬁcant non-linear deformations due to sub-critical damage mechanisms, such as platelet delamination, matrix microcracking, and platelet splitting/fracture, promote strain redistribution and mitigate the intensity of the stress field induced by the crack/notch. This phenomenon is more pronounced for small structures since the size of a fully-developed FPZ is typically a material property and thus its influence on the structural behavior becomes increasingly significant as the structure size is reduced. For sufficiently large structures, the size of the FPZ becomes negligible compared to the structure's characteristic size in agreement with the inherent assumption of the LEFM that non-linear effects are negligible during the fracturing process.

\ii The design of DFC structures must take the foregoing size effect on the nominal strength into serious consideration. In fact, if the correct scaling is not understood and properly acknowledged, the pseudo-ductile behavior reported in fracture tests on small laboratory-scaled DFC specimens may induce designers to overestimate severely the load capacity of real, large structural components. Size effect testing and analysis can overcome this problem.



\ii To characterize the fracture energy, $G_f$, and the effective length of the fracture process zone, $c_f$, an approach combining equivalent fracture mechanics and stochastic finite element modeling was proposed. The model accounts for the effects of the complex random mesostructure of the material by modeling the platelets explicitly. This theoretical framework was able to describe the scaling of structural strength and enabled the characterization of the mode I fracture energy of DFCs. 

\ii $G_f$ and $c_f$ were estimated for a platelet size of $75\times12$ mm, $50\times8$ mm, and $25\times4$ mm respectively. It was found that $G_f=$ $65.47$ N/mm, $55.05$ N/mm, and $33.66$ N/mm while $c_f=$ $13.21$ mm, $6.41$ mm, and $6.54$ mm. These results clearly indicate a strong effect of the platelet size on the fracture properties of the material. In particular, the fracture energy was found to increase linearly with the platelet size in the range investigated in this work. Further computational studies are ongoing to clarify how this trend is affected by the mechanical properties of the constituents and the random orientation of the platelets. 

\ii While the fracture energy increases with the platelet size, the manufacturability indexes \cite{Chang1991} developed for DFCs follow an opposite trend. Accordingly, particular care should be devoted to identifying the platelet morphology providing the best compromise in terms of manufacturabilty and fracture toughness for the desired application.

\ii The analysis of the fracture tests highlighted an outstanding fracture energy of DFCs, from $2.80$ to $5.50$ times larger than the one of a typical Al5083 for the platelet sizes investigated in this work. This result is particularly interesting in view of possible use of DFCs for crashworthiness applications.

\ii Finally, another important conclusion of this work is that, compared to traditional unidirectional composites, DFC structures exhibit higher pseudo-ductility and their strength is, by far, less sensitive to notches, defects, and cracks. However, this aspect can be used advantageously in structural design only upon the condition that proper certification guidelines acknowledging the more pronounced quasibrittleness of DFCs are formulated. The size effect analysis presented in this work represents the first step in this direction as it allows the assessment of the severity of a defect or notch for any DFC structure size and geometry.

 \ee


\section*{Acknowledgments}

This study was financially supported by the FAA-funded Center of Excellence for Advanced Materials in Transport Aircraft Structures (AMTAS) and the Boeing Company. Partial support was also provided by the Joint Center for Aerospace Technology Innovation (JCATI). We thank Ahmet Oztekin, Cindy Ashforth, and Larry Ilcewicz from the FAA, and William Avery from the Boeing Company. We also thank the technical support provided by Bruno Boursier from the Hexcel Corporation, Professor Anthony Waas from University of Michigan and, Bill Kuykendall and Michelle Hickner from the University of Washington. Lastly, we thank all the DFC team members, Kenrick Chan, Reed Hawkins, Rohith Jayaram, Christopher Lynch, Reda El Mamoune, Minh Nguyen, Nicholas Pekhotin, Natania Stokes, Daniel N. Wu, Sam Douglass, and James Davey for their help on the manufacturing and testing.





\clearpage
\listoftables
\listoffigures  
\clearpage


\begin{table}[ht]
\centering  
\begin{adjustbox}{max width=\textwidth}
\begin{tabular}{ c c c c c c} 
 \hline
  \rule{0pt}{4ex}
 Size & Width, $D$  & Gauge length, $L$  & Total length, $L+2L_{tab}^*$  & Notch length, $a_0$  & Thickness, $t$ \\
  & (mm) & (mm) & (mm) & (mm) & (mm)\\[1 ex]
 \hline
$1$ & $120$ & $267$ & $343$ & $24$ & $3.3$ \\
$2$ & $80$ & $178$ & $254$ & $16$ & $3.3$ \\
$3$ & $40$ & $89$ & $165$ & $8$ & $3.3$ \\
$4$ & $20$ & $44.5$ & $120.5$ & $4$ & $3.3$ \\
$5$ & $6.3$ & $14$ & $90$ & $1.3$ & $3.3$ \\

\hline
\multicolumn{4}{l}{$^{\mbox{*}}$ $L_{tab} = 38$ mm}
\end{tabular}
\end{adjustbox}
\caption{\sf Geometry of the Single Edge Notch Tension specimens.}
\label{T1}
\end{table}

\begin{table}[ht]
\centering  
\begin{adjustbox}{max width=\textwidth}
\begin{tabular}{ c c c c c } 
 \hline
  \rule{0pt}{4ex}
 Size  & $75 \times 12$ mm & $50 \times 8$ mm & $25 \times 4$ mm & Quasi-isotropic\\
 \hline
$1$ & $143.2 \pm 16.21$ & $139.1 \pm 15.68$ & $108.9 \pm 9.44$ & -\\
$2$ & $167.7 \pm 27.15$ & $153.8 \pm 27.61$ & $129.3 \pm 0$ & $176.1 \pm 9.73$\\
$3$ & $171.1 \pm 19.09$ & $198.2 \pm 14.90$ & $173.8 \pm 11.10$ & $238.6 \pm 23.20$\\
$4$ & $210.7 \pm 12.58$ & $214.2 \pm 14.58$ & $160.2 \pm 18.87$ & $277.9 \pm 31.54$\\
$5$ & $230.2 \pm 29.13$ & $242.2 \pm 4.99$ & $224.2 \pm 8.29$ & -\\

\hline
\end{tabular}
\end{adjustbox}
\caption{\sf The average failure strength and standard deviations of Single Edge Notch Tension specimens (units: MPa). }
\label{T2}
\end{table}

\begin{table}[ht]
\centering  
\begin{adjustbox}{max width=\textwidth}
\begin{tabular}{ c c c} 
 \hline
  \rule{0pt}{4ex}
 Description &  T$700$G  & Resin   \\ [1 ex]
 \hline
Platelet thickness, $t$ (mm) & $0.135$ & varies \\
In-plane longitudinal modulus, $E_1$ (GPa) & $135$ & $3$ \\
In-plane transverse modulus, $E_2$ (GPa) & $10$ & $3$ \\
In-plane shear modulus, $G_{12}$ (GPa) & $5$ & $1.1$ \\
In-plane Poisson ratio, $\nu_{12}, \nu_{31}$ & $0.3$ & $0.35$ \\

\hline
\end{tabular}
\end{adjustbox}
\caption{\sf Elastic properties of the platelet (T$700$G) and resin layer used in this study.}
\label{T3}
\end{table}

\begin{table}[ht]
\centering  
\begin{adjustbox}{max width=\textwidth}
\begin{tabular}{ c c c c c} 
 \hline
  \rule{0pt}{4ex}
 Platelet size & $g(\alpha_0)^*$  & $g'(\alpha_0)^*$  & Fracture energy, $G_f$  & Effective FPZ length, $c_f$\\
 (mm) &  &  & (N/mm) & (mm)\\[1 ex]
 \hline
$75 \times 12$ & $0.892 \pm 0.058$ & $5.101 \pm 1.579$ & $65.47 \pm 2.81$ & $13.21 \pm 1.85$\\
$50 \times 8$ & $0.852 \pm 0.096$ & $6.607 \pm 2.355$ & $55.05 \pm 6.14$ & $6.41 \pm 0.83$\\
$25 \times 4$ & $0.935 \pm 0.079$ & $5.365 \pm 2.951$ & $33.66 \pm 2.86$ & $6.54 \pm 1.07$\\
Quasi-isotropic & $ 0.84 $ & $4.92 $ & $41.01 $ & $1.85 $ \\
\hline
\multicolumn{4}{l}{$^{\mbox{*}}$ Results of the FEM simulations}
\end{tabular}
\end{adjustbox}
\caption{\sf The fracture properties obtained by the size effect experiments and the stochastic finite element.}
\label{T4}
\end{table}

\clearpage

\end{document}